\documentclass[10pt,conference,twocolumn]{IEEEtran}  
\IEEEoverridecommandlockouts                           
\usepackage{graphics,
			color,
			subfigure,
			calrsfs,
			rotating,
			comment,
			amssymb,
			url,
			epsfig,
			xcolor,
			bm,
			cite,
			textcomp,
			algorithm,
			algorithmic,
			eqparbox,
			amsthm,
			soul}

\usepackage[cmex10]{amsmath}
\usepackage{xspace}
\newtheorem{defin}{Definition}
\newtheorem{lma}{Lemma}

\newtheorem{thm}{Theorem}

\newtheorem{corol}{Corollary}
\usepackage{ulem}

\newcommand\com[1]{%
  \hfill\%{\ #1}%
}

\newcommand {\cofair} {{CoFair}\xspace}
\newcommand {\mps} {{MPS}\xspace}
\newcommand {\1} {{\mathbf 1}}

\newcommand {\C} {{\mathcal C}}
\newcommand {\E} {E}

\newcommand {\Copt}[1]{{C_{#1}^{\mbox{\tiny{OPT}}}}}
\newcommand {\Sopt}[1]{{C_{#1}^{\mbox{\tiny{OPT}}}}}
\newcommand {\Csig}[1]{{\widehat C_{#1}}}
\newcommand {\Ssig}[1]{{\widehat C_{#1}}}
\newcommand {\Clp}[1] {{C_{#1}^{\mbox{\tiny{LP}}}}}
\newcommand {\F}{{\mathcal F}}
\newcommand {\M}{{\mathcal M}}
\newcommand {\Ll}{{\mathcal L}}
\newcommand {\R}{{\mathbb R}}
\newcommand {\St}{{\overline C}}
\newcommand {\Rt}{{\overline R}}

\newcommand {\matlab} {matlab$^{\mbox{\tiny \textregistered}}$\xspace}

\DeclareMathOperator*{\argmin}{\arg\!\min}

\begin{document}
\title{Fair Coflow Scheduling via Controlled Slowdown}
\author{Francesco De Pellegrini$^*$, Vaibhav Kumar Gupta$^{\dagger}$, Rachid El Azouzi$^*$,\\ Serigne Gueye$^*$, Cedric Richier$^*$ and Jeremie Leguay$^\star$\thanks{$^*$Laboratoire informatique d'Avignon (LIA), Avignon University, France, $^{\dagger}$He was with LIA, Avignon University, France and now he is with Interdisciplinary Centre for Security, Reliability and Trust (SnT), University of Luxembourg, $^\star$Huawei Technologies,  Paris Research Center, France.}}
\maketitle
\begin{abstract}
The average coflow completion time (CCT) is the standard performance metric in coflow scheduling. However, standard CCT minimization may introduce unfairness between the data transfer phase of different computing jobs. Thus, while progress guarantees have been introduced in the literature to mitigate this fairness issue, the trade-off between fairness and efficiency of data transfer is hard to control. 

This paper introduces a fairness framework for coflow scheduling based on the concept of slowdown, i.e., the performance loss of a coflow compared to isolation. By controlling the slowdown it is possible to enforce a target coflow progress while minimizing the average CCT. In the proposed framework, the minimum slowdown for a batch of coflows can be determined in polynomial time. By showing the equivalence with Gaussian elimination, slowdown constraints are introduced into primal-dual iterations of the \cofair algorithm. The algorithm extends the class of the $\sigma$-order schedulers to solve the fair coflow scheduling problem in polynomial time. It provides a $4$-approximation of the average CCT w.r.t. an optimal scheduler. Extensive numerical results demonstrate that this approach can trade off average CCT for slowdown more efficiently than existing state of the art schedulers.
\end{abstract}
\pagestyle{plain}


\section{Introduction}\label{sec:intro}


The coflow abstraction has been introduced in the seminal paper \cite{ChowdhuryHotNet2012}. A coflow denotes a group of flows produced by data-intensive computing frameworks, e.g., Map Reduce, Giraph  and Spark \cite{dean2004mapreduce,zaharia2010spark,han2015giraph}, during their data transfer phases. In the same fashion, the workflows of most data-intensive machine learning applications are based on data-transfers and operated by distributing training tasks over hundreds of individual compute nodes. 
A well-studied example of data transfer step is the shuffle phase of Hadoop MapReduce. The shuffle phase acts as a synchronization barrier since the next phase of the computation cannot be pursued until the end of the shuffle data transfer. Once each mapper node finishes running their job, partial results are fetched in the form of HTTP connections by peer reducer nodes and final results can be generated for the next computation phase or storage. At the application level, the shuffle phase has been shown to represent a significant part of the total computation time \cite{chowdhury2015coflow}. For this reason, efficient coflow scheduling has become a key aspect of traffic engineering in modern datacenters \cite{chen2021,chowdhury2019near,agarwal2018sincronia,Mao2018,Utopia2018}. 

The standard metric to measure coflow scheduling performance is the weighted Coflow Completion 
Time (CCT), i.e., the average time by which the last flow of a coflow is completed. Minimizing 
the average CCT is indeed the appropriate goal in order to increase the rate of computing jobs dispatched in a datacenter and improve the execution time of applications. The corresponding minimization problem is complicated by the fact that a shared datacenter fabric can be contended by hundreds of coflows at the same time. Thus, multiple congested links may appear under concurrent demands. In the last ten years, the problem of average CCT minimization has been addressed by several authors \cite{ChowdhuryHotNet2012,chowdhury2015coflow,Shafiee2017Sig,agarwal2018sincronia}, shading 
light on its complexity and devising several algorithmic solutions. The problem is found {\it NP}-hard by reduction to the open-shop problem~\cite{Varys2014}, a mainstream operation research problem where jobs are scheduled on multiple machines. While inapproximability below a factor $2$ has been proved \cite{chowdhury2019near}, to date the best deterministic approximation ratio is $4$ in the case when coflows are released at same time\cite{Shafiee2017Sig,agarwal2018sincronia,Ahmadi2018}. 

In the literature, maximizing network performance is known to entail potential unfairness among different flows \cite{shaksrik008}. Thus, the notion of per-flow fairness has been studied to 
balance the resource allocation, i.e., link utilization, among different flows. Max-min fairness 
and proportional fairness are reference concepts in this context and a series of fundamental works 
on utility-based fairness have showed that both the notions can be compounded under the larger concept of $\alpha$-fairness \cite{Walrand2000}. 

{\it Coflow fairness}. Minimizing the average CCT suffers a similar fairness issue due to starvation \cite{pFabric2013} and fairness has been studied in the context of coflow scheduling as well \cite{LiMaxMin2016}\cite{Boli}\cite{coflex}\cite{chowdhury2016hug}. To obtain a resource allocation able to achieve a required trade-off between average CCT and fairness, it is tempting to directly use CCT as the variable in conventional fairness utility functions. However, a major obstacle to this approach exist. In fact, the definition of CCT entails a minimization, so that using CCT as argument of standard smooth convex functions such as the $\alpha$-fair utility leads to minimization problems neither convex nor differentiable. The workaround is to allow for a constant rate allocation~\cite{chen2018efficient}. But, CCT minimization is a finite horizon problem for which optimizing stationary rates leads to suboptimal resource allocation. 

Fairness in coflow scheduling relies on the notion of {\it coflow progress}, i.e., by guaranteeing a minimum resource allocation to coflows \cite{chowdhury2016hug,Boli,coflex}. In particular, \cite{coflex} studied the trade-off between performance and fairness by showing that it is possible to balance between minimum coflow progress and average CCT. In practice, since coflow links have intertwined dependencies, the coflow progress approach builds on the notion of DRF (Dominant Resource Fairness) \cite{GhodsiDRF} 
in order to allocate rates based on the maximum per port demand. The progress of a coflow tracks the slowest flow -- determining its CCT -- on the related bottleneck port at each instant, while all remaining flows of the coflow are slowed down in order to improve network utilization without impairing the CCT. Finally, a max-min fair argument grants a bottleneck share per coflow in the form of a minimum rate allocation, i.e., a so-called minimum coflow progress guarantee. Apart few exceptions \cite{Utopia2018}, any further performance optimization, e.g., minimization of average CCT or maximization of port utilization, is pursued on top of this baseline constant rate allocation \cite{coflex,chowdhury2016hug,Boli}, denoted as {\it static} progress in the rest of the paper. Note that, from the perspective of a coflow being served, this is just a lower bound on the average rate received before completion.

The notion of coflow fairness based on {\it static} progress has limitations as detailed formally in the next sections. In fact, it may not account for the structure of coflows and their conflicting demands. Furthermore, ensuring a constant minimal allocation corresponds to a non-preemptive definition of progress because it requires every coflow to send a minimum amount of data with no interruption over each engaged port. 

{\it Main contributions.} The fairness framework introduced in this paper is based on a notion of progress which is the average rate granted to a coflow while in service. The coflow progress is tuned by controlling the inverse metric, e.g., the {\it slowdown} \cite{chowdhury2016hug}. The slowdown measures the additional delay on the data transfer time of a coflow compared to isolation, i.e., when a coflow is scheduled alone in the datacenter fabric. An optimal scheduling problem is formulated under a generalized slowdown constraint, obtaining a notion of fairness more general than the notion of progress used in the literature. Similar to $\alpha$-fair scheduling \cite{Walrand2000}, the  {\it slowdown}  constraint acts as a single parameter able to strike the trade-off between efficiency, i.e., CCT performance, and fairness, i.e., the slowdown experienced by a coflow. Furthermore, a fully polynomial time algorithm, namely \mps (Minimum Primal Slowdown) provides an accurate estimation of the minimum feasible slowdown for a batch of coflows. 

Finally, a scheduling algorithm \cofair extends the family of the primal-dual algorithms \cite{mastrolilli,agarwal2018sincronia}: by using the output of \mps, it produces a {\it primal-feasible} $\sigma$-order and an approximation factor of $4$. 

Actually, the target coflow progress is attained based on a per coflow prioritization, avoiding expensive rate-control mechanisms. \cofair is tested against Sincronia \cite{agarwal2018sincronia}, near-optimal for CCT minimization, and against Utopia \cite{Utopia2018} for coflow fairness. \cofair performs closely to Sincronia in CCT and to Utopia in slowdown, thus improving the tradeoff, with low complexity since no rate control is required. To the best of the authors' knowledge, the formal framework for fair scheduling in data-transfer based on controlled slowdown and the solutions provided in this paper are new contributions to the discussion on enforcing fairness in coflow scheduling.

The paper is organized as follows. Sec.~\ref{sec:related} resumes existing works on fair coflow scheduling. Sec.~\ref{sec:model} introduces the system model and the relations between CCT, coflow progress and slowdown. Sec.~\ref{sec:slowdown} describes the mathematical framework as a scheduling problem with slowdown constraints and its feasibility. Algorithmic solutions are proposed in Sec.~\ref{sec:algo} in the set of primal-feasible $\sigma$-order schedulers. Sec.~\ref{sec:numerical} reports on numerical results and a concluding section ends the paper.


\section{Related works}\label{sec:related}


In the coflow literature only a few papers deal with fairness issues. The seminal work~\cite{ChowdhuryHotNet2012} introduced the weighted CCT as objective function and later works pursued same approach, e.g.,~\cite{Zhang2019}, even though the relation between coflow weights, CCT and coflow fairness remains elusive. In general, fairness metrics based either on weights or on constant rate allocation~\cite{Boli,Zhang2019} lead to sub-optimal scheduler and the usage of the CCT geometric mean~\cite{Boli} tends to privilege shorter coflows. In the scheduler Varys~\cite{Varys2014} coflow prioritization has been introduced as a means to mitigate starvation of coflows with low priority. Actually, the smallest-effective-bottleneck-first heuristic used in Varys is a pre-emptive scheduling which prioritizes greedily a coflow with the smallest remaining bottleneck's completion time. 

Thus, granting a constant fraction of port bandwidth per coflow became the accepted notion of coflow progress adopted later on in the literature~\cite{chowdhury2016hug,coflex}. 
This coflow progress guarantee -- already denoted static progress  -- 
has been formalized for the first time in~\cite{chowdhury2016hug}. However, in that context the progress is as a static rate constraint under the general objective of maximizing the fabric utilization. The idea of a progress constraint coupled to the CCT minimization appeared first in \cite{coflex}, where the (static) progress of each flow is to be set above certain target threshold. 

An interesting attempt to develop a formal framework based on max-min fairness has been presented \cite{LiMaxMin2016}, based on lexicographic ordering for flow-level rate allocation \cite{RadunovicTON2007}. However, due to its ease of implementation, the notion of fairness by static coflow progress guarantees has been preferred in the literature. Other authors \cite{qu2019ostb} define the fairness degree of a data-transfer scheme with respect to standard schemes such as Hug~\cite{chowdhury2016hug} or DFR~\cite{GhodsiDRF}. 

More recently, a {\it relative} fairness concept has been introduced in order to compare coflow schedulers to benchmarks such as Hug or DFR~\cite{qu2019ostb}. In particular, \cite{Utopia2018} has proposed a long term isolation guarantee by providing a bound on CCTs relative to DRF performance. While this type of bound resembles the slowdown constraint introduced in this work, the CCT-slowdown tradeoff cannot be controlled. 

In the coflow deadline satisfaction (CDS) problem \cite{Tseng2019,Trung2022} each coflow is subject to a completion deadline similar to a slowdown constraint. However, in that context the target is to operate joint {\it coflow admission control and scheduling} to maximize the number of admitted coflows which respect their deadlines.  

The solution proposed in this paper is rooted on the analysis of a scheduling problem where the minimal slowdown constraint is connected to a primal-dual formulation. The resulting scheduler has no need to perform rate control and it can be fully implemented by greedy rate allocation via priority queuing \cite{agarwal2018sincronia}. The key technical result is the equivalence of the weight adjustment step introduced in \cite{mastrolilli}[Alg. 3.1] -- later applied in \cite{agarwal2018sincronia} to coflow scheduling -- with Gaussian elimination. Coupled with the novel results on primal-feasibility, this leads to an approximation solution based on a $\sigma$-order prioritization algorithm under generalized slowdown constraints. 

In the rest of the paper, reference solutions are near-optimal Sincronia \cite{agarwal2018sincronia} for CCT minimization and Utopia \cite{Utopia2018} for coflow fairness. 


\section{Motivations}\label{sec:model}


\subsection{System model}

\begin{table}[t]
\centering
\begin{tabular}{|p{0.15\columnwidth}|p{0.75\columnwidth}|}
\hline
{\it Symbol} & {\it Meaning}\\
\hline
$\Ll$ & set of switch links $k=1,\ldots,2M$\\
$\M$  & coflow schedule\\
$\C$ & set of coflows $\C=\{1,\ldots,N\}$ \\
$\F_j$ &  set of flows of coflow $j$; $n_j=|\F_j|$ (coflow width)\\
$p_{\ell j}$& volume of coflow $j$ on port $\ell$\\
$C_j$ & coflow completion time for coflow $j$\\
$w_j$ & weight of coflow $j$\\
$r_j$ & release time of coflow $j$\\
$\E$ & slowdown constraint\\
$x_j^i(t)$ & fraction of flow $i$ of coflow $j$ scheduled in slot $t$\\
$Y_j^i(t)$ & progress of flow $i$ of coflow $j$ at time $t$\\
$Z_j(t)$ & completed fraction of coflow $j$ at time $t$\\
$Z_j^i(t)$ & completed fraction of flow $i$ of coflow $j$ at time $t$\\
$x_j^i(t)$ & fraction of flow $i$ of coflow $j$ at  time $t$\\
$X_j(t)$ & indicating variable for coflow $j$ at slot $t$ (binary)\\
$X_j^i(t)$ & indicating variable for for $i$ of. coflow $j$ at slot $t$ (binary)\\
$B_\ell$  &  capacity of link  $\ell$\\
$T$  & time horizon\\
$\Delta$  & time slot  duration\\\hline
\end{tabular}\vspace{1mm}
\caption{List of Key Notations}
\label{notation}
\end{table}

\begin{figure}[t]
\centering
\includegraphics[scale=0.45]{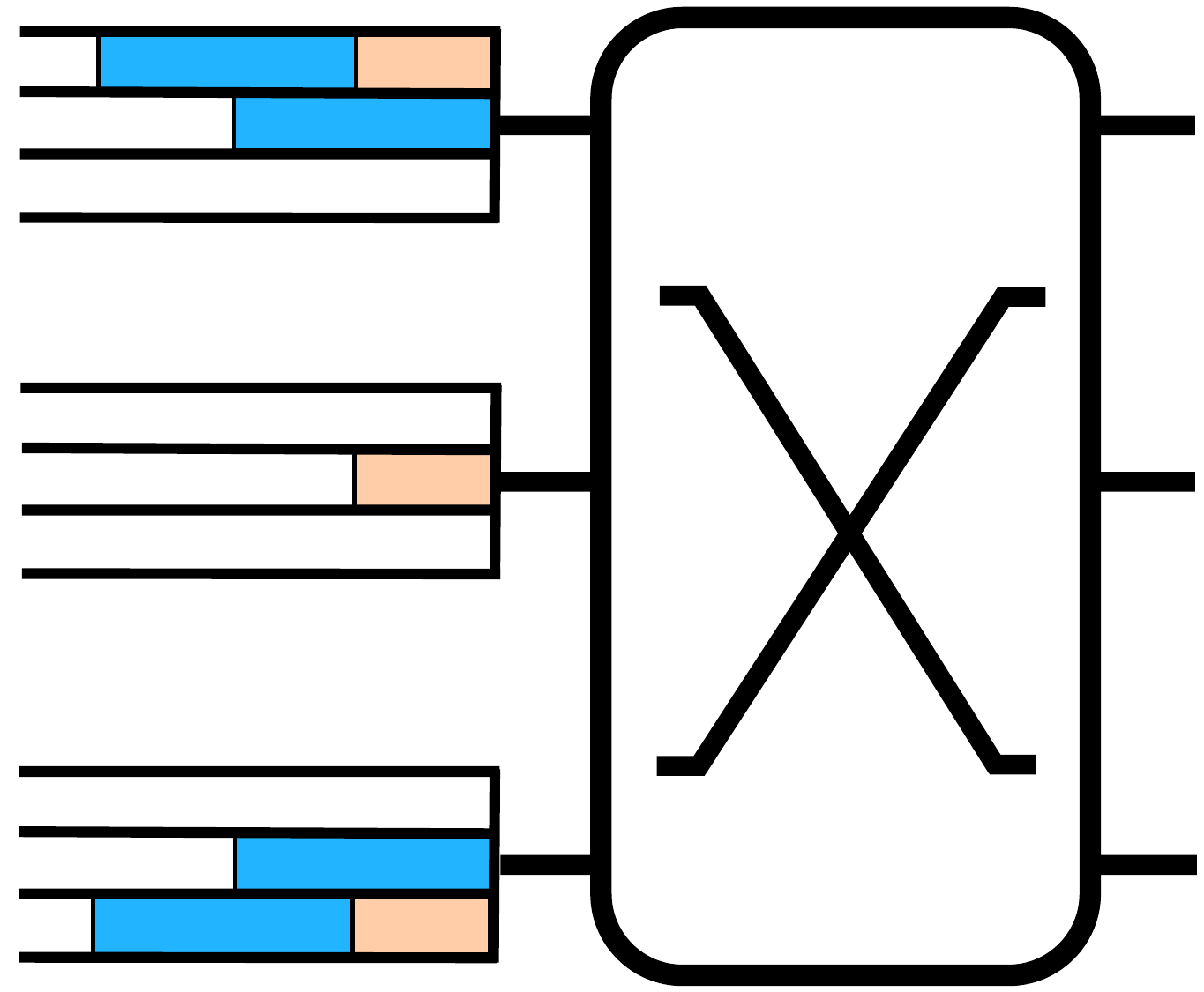} 
\put(-145,126){\put(0,9){\scriptsize V}\put(29,9){\scriptsize  U}\put(18,0){\scriptsize  V}}
\put(-116,73){\scriptsize U}
\put(-145,0){\put(0,7){\scriptsize V}\put(29,7){\scriptsize  U}\put(18,17){\scriptsize  V}}
\put(-100,11){\put(0,12){$\ell_3$}\put(0,68){$\ell_2$}\put(0,121){$\ell_1$}}
\put(-10,11){\put(0,12){$\ell_6$}\put(0,68){$\ell_5$}\put(0,121){$\ell_4$}}
\caption{Coflow scheduling over a $3 \times 3$ datacenter fabric with three ingress and threee egress ports $\Ll=\{\ell_1,\ldots,\ell_6\}$. Flows in ingress ports are organized by destinations and color-coded by coflows: $j=1$ orange and $j=2$ blue.}\label{fig:bigswitch}
\end{figure}
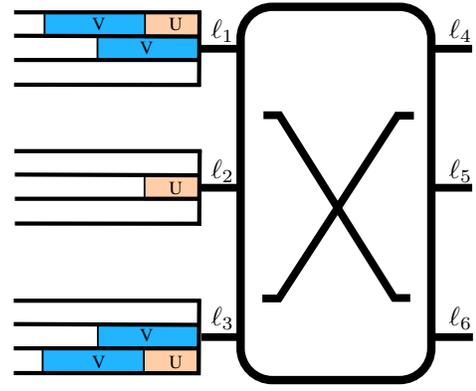

The most popular coflow scheduling model is based on a non blocking switch connection of the type reported in Fig.~\ref{fig:bigswitch}, often called the Big Switch model. This model is adequate because the bisection bandwidth of modern datacenters exceeds access capacity, so that congestion events occur at the inbound or outbound ports of top-of-rack switches. From now on the term link and port will be used interchangeably. The set of switch links is $\Ll=\{1,\ldots,2M\}$, where links $1\leq \ell\leq M$ are ingress ports and $M+1\leq \ell \leq 2M$ are egress ports. Each port has capacity ${B_\ell}$.

Let $\C=\{1,\ldots,N\}$ be an input set of coflows, i.e., a {\it batch}. Each coflow $j$ is a set $\F_j$ of $n_j$ flows and a flow represents a shuffle connection over a pair of input-output ports. The release time $r_j\geq 0$ of coflow $j$ is the time when its shuffle phase starts. A component flow $i\in \F_j$ has volume $v_j^i$. The {\it size} of coflow $j$, that is the total volume of coflow $j$, is denoted by $V_j=\sum_{i \in \F_j} v_j^i$. Let $\chi_j^i(\ell)$ indicate if flow $i\in \F_j$ is active on port $\ell$. The volume of coflow $j$ active on port $\ell$ is denoted
\[
p_{\ell j}=\sum\limits_{i\in \F_j}  v_j^i \chi_j^i(\ell)
\]
Tab.~\ref{notation} summarizes all the important notations and their descriptions.

Let $C_j$ be the CCT of coflow $j$: $C_j$ is the epoch when the last flow in $\F_j$ is fully transferred.  Let $C_j^0$ be the coflow completion time {\it in isolation}, i.e., if all network resources serve solely the tagged coflow. It holds $C_j^0=:r_j+\max\limits_{l \in \Ll} \frac 1{B_\ell}  \sum_{i\in \F_j} v_j^i \chi_j^i(\ell)$. The standard coflow scheduling problem corresponds to determine a coflow schedule $\M$ able to minimize the weighted sum of the CCTs under the capacity constraints imposed by the network fabric. A coflow schedule $\M$ provides the set of rates which are assigned to each flow of each coflow per port. The large class of $\sigma$-order schedulers \cite{chowdhury2019near}, for instance, prescribes to respect a static preemption priority 
$\sigma$ and only requires monitoring of coflows at ingress/egress (I/E) ports of the Big Switch. 
As proved in \cite{agarwal2018sincronia}, the efficiency loss w.r.t. an optimal scheduler respecting the given priority $\sigma$ is at most a factor $2$. In the numerical evaluations in Sec.\ref{sec:numerical}, rates are assigned following a strict priority rule, namely the greedy rate allocation policy \cite{agarwal2018sincronia}. In practice, this rule can be implemented with a priority queuing scheduler with a number of queues equal to the number of coflows. 

\subsection{Static coflow progress} In the coflow literature, the standard definition of fairness~\cite{chowdhury2016hug,Boli} is based on the notion of coflow progress. Given an input batch of coflows, the objective is to minimize the weighted CCT while ensuring that every coflow has a target minimum progress {\it per link}, e.g., per input or output port in the Big Switch model.  Let $a_{\ell j}$ be the rate guaranteed to coflow $j$ on port $\ell$. The progress of coflow $j$ is defined as
 \begin{equation}\label{ea:progress}
P_j = \min_{\ell \in \Ll} \left \{ \frac{a_{\ell j}}{d_{j\ell}}  \right \}
 \end{equation}
where $d_{j\ell}:= \frac{p_{\ell j}}{C_j^0-r_j}$ is also called the rate demand of flow $i$ of coflow $j$ and it represents the lowest possible rate to dispatch $p_{\ell j}$ within $C_j^0$. From the definition of $C_j^0$ indeed $P_j\leq 1$.  Finally, coflow fairness ensures a per-coflow minimum progress \begin{equation}\label{eq:staticfairness} 
P_j \geq P_0, \quad \forall j \in \C
\end{equation} 
It is worth remarking that, though equivalent  to the definition appearing in the literature, \eqref{ea:progress} has a straightforward interpretation in terms of per port {\it bitrates}. In some works, the constant parameter $P_0$ is also called as isolation guarantee \cite{chowdhury2016hug}. The next example is meant to outline why  the notion of static progress does not fully capture the performance-fairness trade-off in coflow scheduling.
\begin{figure}[t]
\hskip3mm\includegraphics[scale=0.65]{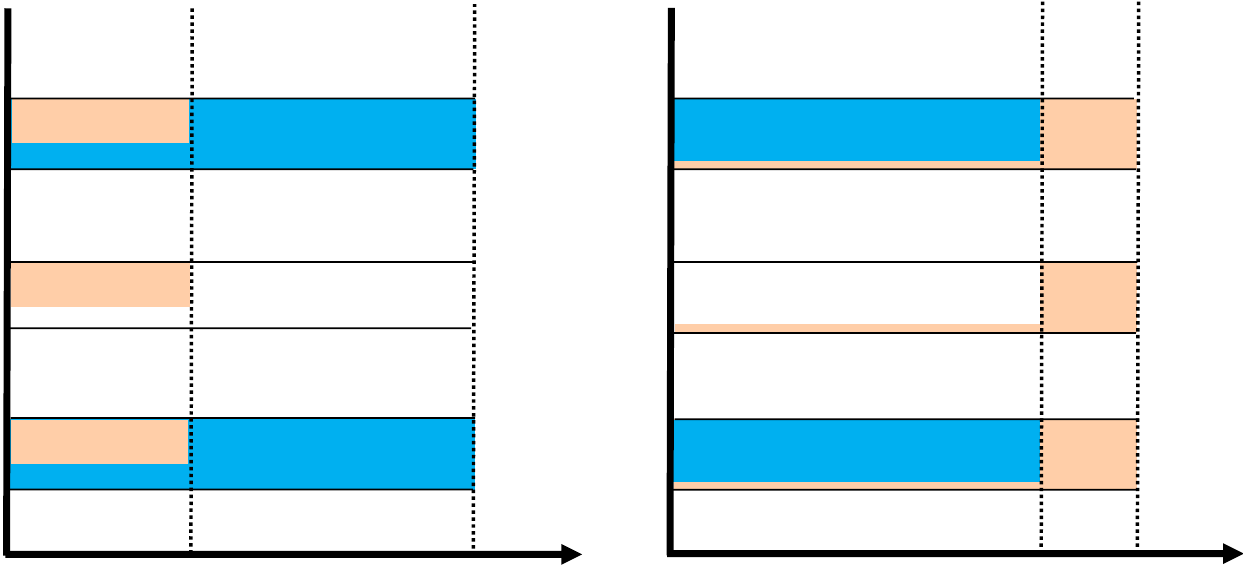} 
\put(-244,0){\put(0,100){a)}\put(0,19){$\ell_3$}\put(0,49){$\ell_2$}\put(0,80){$\ell_1$}\put(10,-7){$\scriptsize 0$}\put(94,-7){\scriptsize $U\!\!+\!\!2V$}\put(37,-7){\scriptsize{$\frac{U}{1-P_0}$}}\put(18,100){\scriptsize{Type I}}}
\put(-120,0){\put(0,100){b)}\put(0,19){$\ell_3$}\put(0,49){$\ell_2$}\put(0,80){$\ell_1$}\put(8,-7){$\scriptsize 0$}\put(95,-7){\scriptsize $U\!\!+\!\!2V$}\put(72,-7){\scriptsize{$\frac{2V}{1-P_0}$}}\put(18,100){\scriptsize{Type II}}}
\vspace{5mm}
\begin{scriptsize}
\hspace*{10mm}\begin{minipage}{10mm}
\begin{tabular}{|c||c|c||c|c|}
\hline
 &  \multicolumn{2}{c}{slowdown ($\phi_j=1$)} & \multicolumn{2}{||c|}{progress} \\ \hline\hline
                          &  $\E_1$  & $\E_2$ & $R_1$ & $R_2$ \\ \hline\hline
       Type I    &  $\frac 1{1-P_0}$  & $1 + \frac{U}{2V}$ & $3(1-P_0)$ & $\frac {4V}{2V+U}$ \\\hline\hline
        Type II    &  $1+ \frac {2V}U$  & $\frac 1{1-P_0}$ & $\frac {3U}{2V+U}$ & $2(1-P_0)$ \\\hline
\end{tabular}
\end{minipage}
\end{scriptsize}
\caption{Example: a) schedule of Type 1 and b) schedule of Type 2.} 
\label{fig:example}
\end{figure}

\subsubsection*{Example} The sample coflow batch of Fig.~\ref{fig:bigswitch} is served on a fabric with $M=3$ servers and unitary bandwidth links. Coflow $1$ marked orange occupies all input and output ports of the fabric; it has $3$ flows with volume $U$ traffic units each. Coflow  $2$ marked blue consists of flows of volume $V$ traffic units each, two on port $\ell_1$ and two on port $\ell_3$. In Fig.~\ref{fig:bigswitch} the index of the virtual queues at ingress ports indicate the output port. For instance the second flow of coflow $2$ sends $v_2^2=V$ traffic units to port $\ell_5$.

Let $P_0$ denote the static progress guarantee. By the symmetry of the problem, same rate guarantee is due on input ports for the two coflows, i.e., $P_1=a_{\ell_1 1}=a_{\ell_21}=a_{\ell_31}\geq P_0$ and $P_2=a_{\ell_1 2} =a_{\ell_3 2}\geq P_0$. Clearly, $P_0\leq 1/2$ due to bandwidth constraints. As depicted in Fig.~\ref{fig:example}, based on the CCT order of coflows, two types of schedules are possible: Type I where coflow $1$ finishes first or Type II where coflow $1$ finishes last. For Type I schedules it holds $C_2=U+V$, whereas $C_1 = U/(1-P_0)$. For Type II, $C_1=U+V$ and $C_2 = 2V/(1-P_0)$. By letting $P_0>0$, for Type I the CCT of coflow $1$ has been degraded with no gain for coflow $2$. Conversely, for Type II schedules, the CCT of coflow $2$ is degraded with no gain for coflow $1$. Thus, while the static progress guarantees a minimum rate allocation per coflow, the CCT of some of the coflows can be degraded without any actual gain for the slowdown of the other ones.  

\subsection{Progress and slowdown} The progress of coflow $j$ under coflow schedule $\M$ is defined as its average rate attained while in the system, i.e., $R_j= \frac{V_j}{C_j-r_j}$. To control the fairness-performance tradeoff, it is convenient to use the coflow {\it slowdown}, which is in fact the inverse quantity of the progress: a target progress guarantee can be enforced via a constraint on the maximum slowdown. The slowdown is the ratio between the progress in isolation $R_j^0$ and the progress under a given schedule $\M$. This plain notion of slowdown can be generalized to allow for a non-negative weight per coflow which is a function of the coflow geometry. 
\begin{defin}[Generalized coflow slowdown]\label{def:gensd}
The generalized slowdown of coflow $j$ under coflow schedule $\M$ is the weighted ratio $\E_j=\phi_j \, \frac{R_j^0}{R_j}=\phi_j \, \frac{C_j-r_j}{C_j^0-r_j}$.  
\end{defin}
The (generalized) slowdown captures the relative degradation of the data transfer duration when a coflow is scheduled within a batch of competing ones compared to its duration in isolation. In the numerical section, two variants of slowdown are considered:
\begin{itemize}
\item {\it plain} slowdown: $\phi_j=1$, this is the standard definition of slowdown \cite{VangCoflex2017,Varys2014,Utopia2018} where, between two coflows, a bound on the maximum slowdown prioritizes the coflow with smaller CCT in isolation; 
\item {\it slowdown with port-occupation}: $\phi_j=V_j$, with this definition, between two coflows with same volume, a bound on the maximum slowdown prioritizes the coflow with smaller port-occupation. 
\end{itemize}
\subsubsection*{Example} The tradeoff between average CCT and progress for Type I and for Type II schedules is described in Fig.~\ref{fig:example}. For Type I schedules the slowdown of the coflow finishing first is $1/(1-P_0)$. The progress guarantee as a function of the static progress guarantee $P_0$ writes $\overline R_{\mbox{\scriptsize {I}}}=\min\{3(1-P_0),\frac {4V}{2V+U}\}$. In case of Type II schedules it writes $\overline R_{\mbox{\scriptsize {II}}}=\min\{2(1-P_0),\frac {3U}{2V+U}\}$. Note  that both functions are non increasing. Also, by letting $P_0=0$, both $\overline R_{\mbox{\scriptsize {I}}}$ and $\overline R_{\mbox{\scriptsize {II}}}$ attain their maximum and the attained average CCT is also minimized for $P_0=0$ for both Type I and Type II schedules. 

Finally, the following cases hold. For $2V<U$ the Type II schedule attains the minimum average CCT, i.e., $(5V+U)/2$ {\it and} the progress guarantee $R^*=3U/(2V+U)$ is also better off than Type I. Conversely for $4V>3U$ a Type I schedule attains the minimum average CCT, i.e., $V+U$ {\it and} the best progress guarantee, i.e., $R^*=4U/(2V+U)$. 

It is interesting to remark that in the above two cases there is {\it no performance-fairness} tradeoff, because minimizing the average CCT is also ensuring the largest minimum progress. 

Conversely, for $2U < 4V < 3U$, a performance-fairness tradeoff does exist: a Type II schedule attains better average CCT but worse progress than a Type I one; on the contrary, a Type I schedule attains worse average CCT but better progress than a Type II one. 

From the example, the both Type I and Type II schedules do attain the best progress guarantee and are better off w.r.t. CCT figures for $P_0=0$ than for $P_0>0$. This fact is expressed as a general result in the next section, ruling out static progress guarantees in fair coflow scheduling.  


\section{Scheduling under slowdown constraints}\label{sec:slowdown}


The standard coflow scheduling optimization framework determines a coflow schedule $\M$ by minimizing the weighted sum of the CCTs under capacity constraints imposed by the network fabric. It is often formulated as a Mixed Integer Linear problem (MILP) for a finite horizon where time is slotted into $T$ time slots of duration $\Delta$; let $\Delta=1$ without loss of generality. The decision variables $\{x_j^i(t)\}$ represent the fraction of flow $i \in \F_j$ of coflow $j$ scheduled in slot $t$. The fraction of volume of flow $i$ of coflow $j$ already transmitted by slot $t$ is  $Z_j^i(t)=\sum_{v=1}^t x_j^i(v)$. The fraction of volume coflow $j$ transmitted by time $t$ writes $Z_j(t):=\frac1{V_j}  \sum\limits_{i\in \F_j} v_j^i Z_j^i(t)$; note that $C_j=\min\{t| \, Z_j(t)=1\}$. Finally, the standard formulation with MILP representing the Progress Scheduling (PS) problem writes
\begin{align}
	\mbox{minimize:}   \; & \sum w_j C_j  \label{eq:dynsched}\tag{PS}\\
	\mbox{subj. to:} \;& \sum_{t=1}^T x_j^i(t)=1, \quad \forall j\in \C, i\in \F_j \label{eq:completion}\\
	&X_j(t) \leq Z_j(t),\;  \forall j\in \C,  \forall t \label{eq:indicating}\\
	&Z_j(t) \leq 0,  \forall t \leq r_j \label{eq:release}\\
	& C_j  \geq 1 + \sum_{z=1}^T (1-X_j(z)) , \;  \forall j\in \C  \label{eq:CCT}  \\
	& \E_j := \frac{C_j - r_j}{C_j^0 - r_j} \leq \frac{\E}{\phi_j}, \; \forall j \in \C \label{eq:elong}\\
	& \sum_{j\in \C} \sum_{i \in \F_i} v_j^i x_j^i(t)\chi_j^i(\ell) \leq B_\ell, \;\forall t, \forall\ell \in \Ll \label{eq:capacity}\\
	& x_j^i(t) \in [0,1],  \; X_j(t) \in \{0,1\}, \;\forall t 
\end{align}
where $E$ is a non-negative parameter and weight $w_j$ denotes the weight of coflow $j$. Let denote $\{C_j^*\}$ a solution of the \ref{eq:dynsched} and $\{R^*\}$ the corresponding progress vector. 

The MILP makes use of coflow completion variables $Z_j(t)$ and binary variables $X_j(t)$. Note that, due to minimization, $X_j(t)$ behaves as the indicating variable of coflow completion, that is $X_j(t)=1$ if $Z_j(t)=1$ and $X_j(t)=0$ if $Z_j(t)<1$. Constraint \eqref{eq:completion} states that each flow should be completed by the end of the time horizon; \eqref{eq:indicating} states that the completion of a coflow is subject to the transfer of the whole coflow volume. In \eqref{eq:release} the release time of coflows imposes the constraint on the beginning of a coflow transmission. Constraint \eqref{eq:CCT} binds coflow completion time and coflow completion variables, which appeared first in \cite{chowdhury2019near}. Finally, \eqref{eq:capacity} provides the constraint on the port capacity; $Z_j(t)$ is defined implicitly in \eqref{eq:indicating} and \eqref{eq:release} for notation's sake. Finally, ~\eqref{eq:elong} is the slowdown constraint to enforce a target coflow progress, where constant $\E\geq \max_j \phi_j$. Note that $\phi_j$ permits to express a preference: for instance, when $\phi_j=V_j$, if two coflows have same total volume, the one which occupies the fabric ports for shorter time is preferred. 

Clearly, adding progress constraints leads to a deterioration of the average CCT with respect to the plain weighted CCT minimization ($E=+\infty$). For the example in the previous section, when $\phi_j=V_j$, the constraint~\eqref{eq:elong} writes $C_1/C_1^0 \leq \E/3U $ and $C_2/C_2^0\leq \E/4V$. Letting $\E=2(U+2V)$ attains the schedule of Type I. The schedule of Type II is attained when $2V<U$. In such case \eqref{eq:elong} is inactive, i.e., there is no tradeoff. However, Type I is to be preferred when the port occupation of coflow $2$ is larger than that of coflow $1$.   

Next, the notion of static progress \eqref{ea:progress} and \eqref{eq:staticfairness} existing in the literature \cite{coflex} and that of progress based on coflow slowdown are compared. The Static Progress Scheduling (SPS) problem is described below:
\begin{align}
	\mbox{minimize:}   \; & \sum w_j C_j  \label{eq:botsched}\tag{SPS}\\
	\mbox{subj. to:} \;&  \eqref{eq:completion}\eqref{eq:indicating}\eqref{eq:release}\eqref{eq:CCT}\eqref{eq:capacity}\\	
	\;& \sum\limits_{i\in \F_j} v_j^i x_j^i(t) \chi_j^i(\ell) \geq a_{\ell j} \, \Delta\, ,  \;\forall t \; \forall \ell \in \Ll  \label{eq:staticfair}\\	
	\;& x_j^i(t) \in [0,1],  \; X_j(t) \in \{0,1\}, \;\forall t 
\end{align}
where the minimum rate allocations $\{a_{\ell j} \}$ obey to \eqref{ea:progress} and \eqref{eq:staticfairness} for a given progress parameter $P_0$. With some abuse of terminology, let denote $\{\St_j\}$ a solution of \ref{eq:botsched}. 

From an analogous result for the weighted CCT minimization, it follows that both \ref{eq:dynsched} and \ref{eq:botsched} problems are NP-hard and inapproximable below a factor $2$ \cite{chowdhury2019near}.

Let $\Rt_j$ the average progress guarantee offered by \ref{eq:botsched} for coflow $j$ corresponding to allocation $\{a_{\ell j}\}$ so that $\Rt_j :=\frac 12 \sum_{\ell} \sum_{i \in \F_j}a_j^i \; B_{\ell}$. As shown next, compared to \eqref{eq:staticfair}, the slowdown constraint \eqref{eq:elong} allows more flexible rate allocation since the rate on some links can be increased by reducing or pausing the transmission on some other links used by the coflow. Formally, for every \ref{eq:botsched} scheduler there exists a slowdown value $\E$ and a \ref{eq:dynsched} scheduler which is better off.
\begin{thm}\label{thm:dyn}
Let consider $\C$ and static progress defined for $P_0>0$ according to \eqref{ea:progress} and \eqref{eq:staticfairness}. There exists $\E\geq 0$ such that a solution of \ref{eq:botsched} is feasible for \ref{eq:dynsched}. Furthermore, there exists a schedule such that 
\[
\sum C_j^* < \sum \St_j, \quad \mbox{and} \quad R_j^* \geq  {\Rt}_j \quad  \forall  j\in \C
\]
\end{thm}
\begin{IEEEproof}
i. Let $\M$ be a schedule solving \ref{eq:botsched} for a given pair $\{a_{j\ell}\}$ and $P_0$.  Let $\{\St_j\}$ the corresponding result. It is possible to determine $\E$ so that such a schedule is a solution of \ref{eq:dynsched} as well. From \eqref{ea:progress} and \eqref{eq:staticfairness}
\[
\sum\limits_{i\in \F_j} v_j^i x_j^i(t) \chi_j^i(\ell) \geq a_{j\ell} \, \Delta, \quad \quad  \frac{ a_{\ell j} \; (C_j^0 - r_j)}{p_{\ell j}} \geq P_0
\]
For every coflow we can write 
\begin{eqnarray}\label{eq:constrE}
&&\hskip-5mm \St_j -r_j \leq \max_{\ell \in \Ll}  \left \{ \frac{p_{\ell j}} {a_{\ell j} } \right \}= \frac{ (C_j^0-r_j)}{\min\limits_{\ell \in \Ll}\frac{ {a_{\ell j}} \;(C_j^0-r_j)}{p_{\ell j}}} \nonumber\\
&&\hskip-5mm = \frac{C_j^0-r_j}{P_j} \leq   \frac{C_j^0-r_j}{P_0}, \qquad\forall j \in \C  
\end{eqnarray}
With the identification $\E = \frac {\max V_j}{P_0}$, the schedule $\M$ is feasible also for \ref{eq:dynsched}. \\%
ii. The attained progress guarantees can be bounded by
\begin{eqnarray}\label{eq:proguar}
&&\hskip-5mm\Rt_j = \frac1{2}\sum_{\ell \in \Ll} a_{\ell j} \; B_{\ell}  \leq  \frac12 \frac1{\St_j-r_j} \sum\limits_{t=r_j}^{\St_j}  \sum_{\ell \in \Ll} \sum_{i \in \F_j} v_j^i x_j^i(t) \chi_j^i(\ell)\nonumber\\
&& = \frac12  \frac{ 2V_j}{\St_j-r_j}  = R_j
\end{eqnarray}
where the factor $2$ appears since every flow is counted at the ingress at the egress port. By adding feasible constraints $C_j - r_j\leq V_j/\Rt_j$ to  \ref{eq:botsched} 
and dropping \eqref{eq:staticfair} the thesis follows. 
\end{IEEEproof}
From Thm.~\ref{thm:dyn}, scheduling with slowdown constraints provides larger feasibility space and at least same progress guarantees than \ref{eq:botsched}. As seen in our example, the corresponding minimum weighted CCT which can be attained under same progress guarantees is strictly worse for $P_0>0$. However, the introduction of slowdown constraints on top of weighted CCT minimization poses the issue of feasibility, i.e., whether or not, for a given instance of the problem, a target slowdown constraint is feasible and what is the smallest feasible value of $E$. 

\section{Slowdown feasibility}

In this section the feasibility of \ref{eq:dynsched} is discussed and means to test it in polynomial time are presented.  In particular, the largest possible average progress is obtained by minimizing $\E$ in \eqref{eq:elong} while ensuring the feasibility of \eqref{eq:dynsched}. Then, the class of $\sigma$-order preserving schedulers is extended to account for slowdown constraints.

Testing slowdown feasibility is possible in the form of a linear program (LP) feasibility test, from which the following result hold: 
\begin{thm}\label{thm:feasib}
i. The feasibility of \ref{eq:dynsched} can be determined in polynomial time. ii. If an instance of \ref{eq:dynsched} is feasible for a given value if $\E$ where $r_j=0$ for all $j\in \C$, then it is also feasible for all instances obtained by letting some of the $r_j$s positive. 
\end{thm}
The proof and the actual LP formulation are detailed in the Appendix. We denote $\E^*$ the minimum feasible parameter appearing in \eqref{eq:elong}. It is interesting to remark that from the previous result the case of different release times is beneficial with respect to the coflow progress; the intuition is that for a later released coflow, it is convenient if some of earlier coflows have been already dispatched. In the next section we shall introduce a much faster approximated yet accurate algorithm to determine $\E^*$ (namely Alg.~\ref{min_w}). From Thm.~\ref{thm:feasib}, $\E^*$ can be determined in pseudo-polynomial time by solving LP iteratively with a bisection search exploring the slowdown parameter $\E$. By simple calculations this search is $O(\frac {(NV)^{3.5}}\epsilon)$ where $V=\sum V_j$, and $\epsilon>0$ is a tolerance, where we used the $O(n^{3.5})$ bound in the number of variables $n$. 

\subsection{$\sigma$-order schedulers} Even if the above result is encouraging, one fundamental problem with techniques based on MILPs in coflow scheduling is that the number of flows per coflow maybe large\footnote{It may range in the order of tenths of thousands~\cite{chowdhury2015coflow} in production datacenters}. Furthermore, the number of variables involved in such time-indexed systems increases with the inverse of the time slot $\Delta$. Hence, the number of per slot rate decision variables easily range in hundred of thousands for a time slot $\Delta$ in the order of seconds. Thus, scalability issues do arise even for approximation algorithms obtained by solving LPs from MILP relaxation \cite{chowdhury2019near},\cite{Shafiee2017Sig}. Same scalability issues occur solving the time-indexed LP to determine the minimal slowdown $\E^*$.  

For the sake of striking a fairness-performance tradeoff, solving \ref{eq:dynsched} exactly may not be feasible. Rather, it is possible to design a coflow scheduler to approximate a target average progress guarantee, and yet perform near-optimally with respect to the weighted CCT. We can regard this approach as a {\it soft} coflow fairness guarantee versus the  {\it hard} coflow fairness guarantee provided by \ref{eq:dynsched}. 

It is possible to do so using the class $\Sigma$ of $\sigma$-order preserving schedulers ($\sigma$-order schedulers for short). These schedulers require to set the priority of coflows and use such order to determ rate allocation \cite{agarwal2018sincronia,Ahmadi2018}. Let order $\sigma: \C \rightarrow \C$ be a permutation of the coflow set: $j=\sigma(k)$ means that the $k$-th priority is assigned to coflow $j$. 
\begin{defin}
Fix an order $\sigma: \C \rightarrow \C$: the scheduler $\M$ is a $\sigma$-order preserving scheduler if
\begin{itemize}
\item it is pre-emptive;
\item it is work conserving;
\item it respects the pre-emption order: no flow of coflow $\sigma(j)$ can be pre-empted by a flow of $\sigma(k)$, $k=j+1,\ldots N$ on any port on which it has pending flows.
\end{itemize}
\end{defin}
The greedy rate allocation, for instance, blocks a coflow $\sigma(j)$ on some port iff it is occupied by some coflow $\sigma(k)$, $k<j$. As first proved in~\cite{agarwal2018sincronia}, the design of $\sigma$-order schedulers can be performed using the analogy of the Big Switch ports with correlated machines and using related results for open-shop scheduling \cite{mastrolilli}. Following the same idea, once $E$ is fixed, schedulers in $\Sigma$ can approximate the solution of \ref{eq:dynsched} problem. In the rest of the discussion, we consider the case $r_j =0$, $\forall j \in \C$.

\subsection{Extending $\sigma$-order schedulers: primal feasibility} The theory of $\sigma$-order scheduling is rooted on the work \cite{mastrolilli} for concurrent open shop problems where jobs are scheduled on parallel machines. Let define $p_{\ell j}=\sum v_j^i \chi_j^i(\ell)$ the total volume to be transferred by coflow $j$ over port $\ell$. Denote $V_{\ell}=\sum v_j^i \chi_j^i(\ell)$ the total volume on port ${\ell}$ and  $\C_{\ell}$ is the set of coflows active on port $\ell$. The equivalent slowdown constraint $D_j=\frac{\E C_j^0}{V_j}$ is introduced for notation's sake. With the job scheduling terminology, $p_{\ell  j}$ is the duration of the job $j$ over machine $\ell$. The corresponding scheduling formulation writes \cite{mastrolilli}
\begin{align}
	\mbox{minimize:}   \; & \sum_{j\in \C} w_j C_j  \label{eq:primal}\tag{Primal LP}\\
	\mbox{subj. to:} \;& \sum_{j \in A} p_{\ell j} C_j \geq f_{\ell}(A), \quad \forall A \subseteq \C, \; \ell \in \Ll \label{eq:parallel}\\
	& C_j   \leq D_j ,  \; \forall j \in \C  \label{eq:elongLP}
\end{align}
The set functions $f_{\ell}:\mathcal P(\C_\ell) \rightarrow \R$ are defined $f_{\ell}(S) = \frac12 [( \sum_{j\in S} p_{\ell j} )^2 + \sum_{j\in S}  p_{\ell j}^2 ]$ for $S\subset \C_\ell$.  Constraints \eqref{eq:parallel} are known as parallel inequalities. 
Let restrict to the set of schedulers $\Sigma_b$, that is the set of $\sigma$-order schedulers such that the permutation $\sigma$ is {\it primal-feasible}, that is 
\begin{equation}\label{eq:prim-fea}
C_{k\ell} := \frac1{B_\ell}  \sum_{j=1}^k p_{\ell\sigma(j)} \leq D_k, \qquad  \forall \ell \in \Ll, \forall k \in \C
\end{equation}
In fact, it is always possible to consider just $\sigma$-order schedulers
\begin{thm}\label{thm:sigmafea}
If \ref{eq:primal} is feasible, it is solved by $\sigma$-order which is primal-feasible.
\end{thm}
\begin{IEEEproof}
Let consider the set of permutation schedules $\{\sigma_\ell\}$ which schedule coflows active on port $\ell$  according to the EDD rule (Earliest Due Date rule) \cite{mh2021}. Since \ref{eq:primal} is primal-feasible, all such schedules $\{\sigma_\ell\}$ are feasible. Now let derive a feasible $\sigma$-order using the following procedure: select the coflow $k$ for which $C_k$ is maximum, and modify  $\sigma_\ell$s for all ports where $p_{\ell k}>0$ by scheduling $k$ last on all its active ports, and leaving the order of the others unchanged. After this operation it holds $C_{k\ell}\leq D_k$ and $C_{j\ell}\leq D_j$ for all $j\not = k$ and for all ports $\ell$ so that the resulting $\{\sigma_\ell\}$ are still feasible. Let $\sigma(k)=|\C|$, and then remove it from $\{\sigma_\ell\}$. Let $\C\leftarrow \C\setminus \{k\}$. By repeating this procedure, in $N$ steps a feasible $\sigma$-order is obtained.\end{IEEEproof}
Let $\E^p$ be the minimal value such that \ref{eq:primal} is feasible. From the proof of Thm.~\ref{thm:primalfea}, a bisection search algorithm based on a per port EDD scheduling can determine the value $\E^p$ in pseudo-polynomial time. The next result provides a connection between the \ref{eq:primal} and \ref{eq:dynsched}:
\begin{thm}\label{thm:primalfea}
If \ref{eq:dynsched} is feasible, then \ref{eq:primal} is feasible, i.e., $\Sigma \subseteq \Sigma_b$. 
\end{thm}
\begin{IEEEproof}
It is sufficient to consider a schedule $\M$ which is feasible for \ref{eq:dynsched} and then consider the $\sigma$-order which sorts coflows according to the CCTs produced by $\M$, denoted $\{\overline C_k\}$. In fact, let consider coflow $k\in \C$ and port $\ell$ where the last of its flows completes. Since coflow $k$ completes on port $\ell$ after all coflows $j=1,\ldots,k-1$, the volume of all flows on $\ell$ which belong to $j=1,\ldots,k-1$ must be over by time $\overline C_k$. This implies that 
\begin{equation}
D_k \geq \overline C_{k} \geq \sum_{j=1}^k  p_{\ell j}
\end{equation}
so that \eqref{eq:elongLP} holds.
Moreover, it holds 
\[
\sum_{j\in \C}  p_{\ell,j} \overline C_{j} \geq \sum_{j\in \C}  p_{\ell,j} \sum_{k=1}^j  p_{\ell k}=f_\ell(\C)
\]
so that \eqref{eq:parallel} is satisfied as well. By identifying $C_{j}:=\sum_{k=1}^j  p_{\ell k}$, the proof holds. 
\end{IEEEproof}
Thus, if there exists a solution for \ref{eq:dynsched} in $\Sigma$, then it is to be found in  $\Sigma_b$. However, since primal-feasibility does not account for cross-dependencies among ports, it provides a weaker estimate of the completion time for the flows of a coflow according to a given $\sigma$-order. 
\begin{corol}\label{thm:lowerbound}
$\E^p\leq \E^*$.
\end{corol}
Nevertheless, the experimental results reported in Sec~\ref{sec:numerical} indicate that the two values practically coincide. To this respect, the calculation of $\E^p$ provides a substantial computational advantage since it can be performed by a lightweight {\it fully polynomial time algorithm}. It is called \mps, which stands for Minimum Primal Slowdown. 

The pseudocode of \mps is reported in Alg.~\ref{min_w}.  
\begin{algorithm}[t]
	\caption{Pseudocode of \texttt{MPS}}\label{min_w}
	\begin{small}
		\begin{algorithmic}
			\STATE    \textbf{Input:}   ${\cal C}$, $\{\phi_j\}$  $\{p_{\ell,j}\}$ 
			\STATE    \textbf{Output:}  $\E$		
			\end{algorithmic}
		\begin{algorithmic}[1]
		        	\STATE Sort $\C$ in decreasing order of ${\widetilde R}_j^0=\frac{\phi_j}{V_j} \frac{V_j}{C_j^0}$ \com{sorting by $\mathcal R_j^0$} 
			\STATE    $\E_{\ell} = +\infty, \quad \forall \ell  \in \Ll$      		  \com{initialise slowdown}		
			\FOR{$\ell=\ell_1,\dots,\ell_{2M}$} 
			\FOR{$j \in \C_{\ell}$}
				\STATE $Z_j = {\widetilde R}_j^0 \sum_{k=1}^{j} v_k^{\ell} $ 	\com{per coflow estimate on $\ell$} 
			\ENDFOR
		    \STATE $\E_{\ell}= \max\{Z_1,...,Z_{|\C_{\ell}|}\}$ \com{slowdown estimate per port }   
		        \ENDFOR
		    \STATE $\E = \max\{\E_{\ell_1},\ldots,\E_{\ell_{2M}}\}$		\com{for every port}  
		    \RETURN $\E$ \com{return the slowdown} 
		\end{algorithmic}
	\end{small}
\end{algorithm}
The value of $\E$ which satisfies the slowdown constraint \eqref{eq:elong} is minimum when the CCT of the coflow with the largest value $\frac{\phi_j C_j}{C_j^0}$ is minimized. The algorithm visits each port $\ell$ by sorting $\C_{\ell}$ in decreasing order of ${\widetilde R}_j^0=\frac{\phi_j}{V_j} \frac{V_j}{C_j^0}=\frac{\phi_j}{V_j} R_j^0$. For each link $\ell$, the iteration at step $7$ provides a lower bound on $\E^p$ per coflow active on $\ell$ using~\eqref{eq:elong}. The maximum among such estimates is selected at step $9$. Due to the sorting at line $1$ and the $N_\ell$ products at line $5$, it is easy to see that the complexity of Alg.~\ref{min_w} is in $O(MN + N\log(N))$. The next result proves that the output of \texttt{MPS} is correct.
\begin{thm}\label{thm:vbalgo}
\texttt{MPS} returns $E^p$
\end{thm}
\begin{IEEEproof}
From \ref{eq:primal}, fixed a primal schedule, it holds $E \geq {\widetilde R}_j^0 \,C_j$ for $\forall j\in \C$, and more precisely
\[
E^p = \min\limits_{\sigma \in \Sigma_b}\left \{  \max\limits_{\ell \in \Ll}  \max\limits_{j \in \C}  {\widetilde R}_j^0 \, \sum\limits_{i\in \C} p_{\ell \sigma(i)} \right \}
\]
Let show that $E^p$ is attained by a primal schedule which sorts coflows with decreasing ${\widetilde R}_j^0$. Let assume ${\widetilde R}_1^0\geq \ldots \geq {\widetilde R}_N^0$. Let $\sigma_\ell$ be the permutation induced by $\sigma$ on a tagged port $\ell \in \Ll$. For the last scheduled coflow on port $\ell$ it holds
\begin{equation} 
E^p \geq \min\limits_{j \in \C}  {\widetilde R}_j^0 \, \sum\limits_{i\in \C} p_{\ell \sigma(i)}  = {\widetilde R}_{N_\ell}^0  \sum_{i=1}^{N_\ell} p_{\ell i} 
\end{equation}
Also, let $\sigma'=(\sigma(1),\ldots,\sigma(N_\ell-1))$, it holds  
\begin{equation} 
E^p \geq  \max_{j \in  \C_\ell \setminus \{N_\ell \}} \max \left \{ {\widetilde R}_j^0 \sum\limits_{\sigma'(i) \leq \sigma'(j)} p_{\ell \sigma'(i)}  \right \}  \nonumber \\
\end{equation}
because by eliminating the last scheduled coflow $\sigma(N)$, the maximum slowdown cannot increase. Now, it is possible to write
\begin{eqnarray}
&& \hskip-5mm E^p \geq \max\limits_{\sigma_\ell(N_\ell)} \left \{  {\widetilde R}_{\sigma(N_\ell)}^0  \sum_{i=1}^{N_\ell} p_{\ell i},  \max\limits_{j \in \C_\ell \setminus \{N_\ell\} } a_j \sum\limits_{\stackrel{i \in \C_\ell \setminus \{N_\ell\}}{\sigma(i)\leq \sigma'(j)}} p_{\ell \sigma'(i)} \right \} \nonumber \\
&& \hskip-5mm \geq \max \left \{ {\widetilde R}_{N_\ell}^0  \sum_{i=1}^{N_\ell} p_{\ell i},  \max_{j \in  \C_\ell \setminus \{N_\ell \}}\sum\limits_{\sigma'(i) \leq \sigma'(j)} p_{\ell \sigma'(i)} \right \} \nonumber 
\end{eqnarray}
Hence an optimal permutation writes $\sigma=(\sigma_1,\ldots,\sigma_{N_\ell-1},N_\ell)$. The same argument holds for $\sigma'$ and $\C_\ell \setminus \{N_\ell \}$, so that an optimal permutation writes $\sigma=(\sigma_1,\ldots,\sigma_{N_\ell-2}, ,N_{\ell-1},N_\ell)$. After $N_\ell$ iterations, the theorem follows.  
\end{IEEEproof}
As a side result of the above proof it follows
\begin{corol}\label{thm:feascrit}
 \ref{eq:primal} is feasible if and only if the schedule which sorts coflows with decreasing ${\widetilde R}_j^0$ is feasible.
\end{corol}

\subsection*{On feasibility and primal feasibility.} Finally, the minimum value of the parameter $E^p$ for which the formulation is primal-feasible is in general smaller than the minimum possible value $\E^*$ which renders \ref{eq:dynsched} feasible. The fact that they practically coincide -- as seen in the numerical section -- is to be ascribed to the fact that the largest slowdown corresponds invariably to the CCT of a coflow finishing last on some port. Hence, it is in fact determined by the total volume transferred on that port. This is actually the key step for the slowdown estimation performed by \mps at line (5).  

A primal-feasible $\sigma$-order may not grant the target slowdown in \ref{eq:dynsched} for all coflows once rate allocation is performed. However, the objective is to provide guarantees on the slowdown of coflows by using primal-feasibility as a soft constraint. To this respect the feasibility of \ref{eq:dynsched} can be regarded as a hard constraint on the slowdown. In all numerical experiments described in Sec.~\ref{sec:numerical} the fraction of violations never exceeded a few percents.


\section{Algorithmic solution}\label{sec:algo}


This section introduces a primal-dual algorithm, namely \cofair, which can determine a primal-feasible $\sigma$-order in polynomial time. It generalizes the primal-dual scheduling approach \cite{mastrolilli} for job scheduling over uncorrelated machines, a technique later applied in the context of coflow scheduling in \cite{agarwal2018sincronia} and \cite{Ahmadi2018}. The basic idea in \cite{mastrolilli} is to utilize the Smith rule over different machines and sort jobs accordingly: machines are mapped into ports, jobs into coflows and tasks into flows. Every port is treated as uncorrelated to the others and a primal-dual iteration performs a coflow selection and weigh-adjustment technique \cite{mastrolilli,agarwal2018sincronia}

Nevertheless, it is not obvious whether or not it is possible to perform a similar primal-dual iteration while accounting for slowdown constraints and ultimately obtain a primal-feasible schedule. The rest of the section shows how to generate a primal-feasible $\sigma$-order if one exists. The pseudocode of the algorithm described in this section requires specific notation which are introduced next. 

For the set of coflows $A\subseteq \C$, let $V_{\ell}(A)=\sum v_j^i \chi_j^i(\ell)$ the total volume transmitted over port $\ell$; if $A=\C$ the notation is simply $V_{\ell}$. The set of ports engaged by coflows in $A$ is $\Ll(A)=\{\ell \in \Ll | V_{\ell}(A) > 0 \}$. $\C_{\ell}(A)$ is the set of coflows in $A$ active over port $\ell$. $T_{\ell}(A) = \frac1{B_\ell} V_{\ell}(A)$ is the minimum time required to complete the last coflow of set $A$ over link $\ell$. Coflow $j$ is a feasible {\it tail coflow} of set $A$ if $T_{\ell}\leq D_j$ for each link $\ell$ where $p_{\ell,j}>0$. Formally, $F(A)=\{j \in \C_{\ell}(A)| T_{\ell}(A) \leq D_j, \; \forall\ell \in \Ll(A) \}$ is the set of feasible {\it tail coflows} of set $A$. With no loss of generality, in the rest of this section $B_\ell=1$. 

\begin{algorithm}[t]
	\caption{Pseudocode of \cofair}\label{alg:greedy}
	\begin{small}
		\begin{algorithmic}[1]		
		        \STATE  \textbf{Input:}    ${\cal C}$, $\{p_{\ell,j}\}$, $\{D_j\}$, $\{\alpha_j\}$
                         \STATE  \textbf{Output:}  $\sigma$ $\{y_{_{\mu_k,F_k}}\}$ $\{C_j\}$
                         \STATE $\{y_{\ell,S}\}=0$ for $S\subseteq \C$            \com{initialise the dual variables}
                         \STATE  $A_N = \C$	         				         \com{initial unscheduled coflows} 
                         \STATE  $w_j^{(N+1)}= w_j +\alpha_j$			 \com{initial weights}
			\WHILE {$A_k \not = \emptyset$}   				
			\STATE  $\mu_k  \leftarrow$ \mbox{pivot\_bottleneck}$(A_k)$ \com{pivot bottleneck}
			\STATE  $F_k \leftarrow  F(A_k)$ \com{set of feasible tail coflows in $A_k$}
            \IF{$\F_k=\emptyset$}
                \RETURN $\emptyset$ \com{\ref{eq:primal} non-feasible }
            \ENDIF	
			\STATE  $\sigma(k) \leftarrow \arg\min_{j \in F_k} \left \{\frac{w_j^{(k)}}{p_{\mu_{k,j}}} \right \}$ \% pivot coflow (Smith rule)	
			\STATE  $w_{\sigma(k)}^{(k)} \leftarrow \frac{w_{\sigma(k)}^{(k+1)}}{p_{\mu_{k},j}}$  \com{update pivot coflow weight}
		         \STATE   $y_{{\mu_k},F_k} \leftarrow {w_k}^{(k)}$  \com{update dual problem solution}
			\FOR{$j \in A_k \setminus \{\sigma(k)\}$}     			
			         \IF{$j \in \F_k$}
			         \STATE  $w_j^{(k)} \leftarrow w_j^{(k+1)} - w_{\sigma(k)}^{(k+1)}\frac{p_{\mu_{k,j}}}{p_{\mu_{k},\sigma(k)}}$  \com{update weights}
			         \ELSE	 
			         \STATE  $w_j^{(k)} \leftarrow w_{j}^{(k+1)}$  \com{non tail-feasible coflows: unchanged} 
				 \ENDIF				 
			\ENDFOR
			\STATE  $A_{k-1} \leftarrow A_k\setminus \{\sigma(k)\}$	\com{update unscheduled coflow set} 		
		        \ENDWHILE
		        \STATE   $C_j:=\max_{\ell\in \Ll} \sum_{h=1}^j p_{\ell,\sigma(h)}, \quad j\in \C$ \com{CCT lower bound} 		
		        \RETURN $\sigma$, $\{y_{_{\mu_k,F_k}}\}$, $\{C_k\}$ 
		\end{algorithmic}
	\end{small}
\end{algorithm}
The \cofair pseudocode in Alg.~\ref{alg:greedy} describes the primal-dual iterations to generate a $\sigma$-order which is primal-feasible. At each step $k$ it tries to identify a {\it pivot} bottleneck-coflow pair (line $7$ and $12$). When this is not possible, the algorithm exits with returning a non-feasibility result (line $10$). Otherwise, first, a target bottleneck port $\mu_k$ is returned for the unscheduled coflows $A_k=\{\sigma(1),\sigma(2),\ldots,\sigma(k)\}$ using a {pivot\_bottleneck} procedure (line $7$). Second, it identifies the coflow $\sigma(k)$ to be scheduled {\it last} over such port in the set of feasible tail coflows $F_k$. On the selected bottleneck, a feasible tail coflow is chosen according to the Smith rule \cite{queyranne}. Weight updates are only performed for the pivot coflow (line $8$) and for feasible tail coflows (line $12$). Afterwards, the algorithm eliminates the selected coflow from the set of unscheduled ones (line $17$). Weights $\alpha_j$ have the meaning of multipliers and can be used in order to enforce different $\sigma$-orders: as it will be showed in the next section using such weights and {pivot\_bottleneck} it is possible to attain all possible primal-feasible $\sigma$-orders. 

The {pivot\_bottleneck} procedure can be fully general, but in the numerical tests the selection is based on the heuristics that selects the most charged bottleneck. This is the baseline rule adopted in \cite{mastrolilli,Ahmadi2018,agarwal2018sincronia}. Direct calculations show that the computational complexity of \cofair is $O(N(M+N))$. 

\subsection{Correctness}
We prove first that \cofair terminates in $N$ steps: at each iteration it is possible to find at least one feasible tail coflow to be selected at step (8).
\begin{lma}\label{lem:feas}
If \ref{eq:primal} is feasible, then $F_k\not = \emptyset$ for $k=1,\ldots,N$.
\end{lma}
\begin{IEEEproof}
For $k=N$ the statement is true: $A_N=\C\not = \emptyset$ and $F_N\not=\emptyset$, otherwise \ref{eq:primal} would not be feasible. Hence, Cor. \ref{thm:feascrit} ensures that the $\sigma$-order which sorts coflows in order of decreasing values of $R_j^0$ is feasible. From the proof of Thm. \ref{thm:vbalgo} such schedule is feasible 
for any $A_k\subseteq \C$ so that $F_k\not =\emptyset$ for all $k=1,\ldots,N$. 
\end{IEEEproof}
From the proof of Lemma \ref{lem:feas}, we can observe two facts: i) the statement is true irrespective of the implementation of the {pivot\_bottleneck} procedure; ii) it holds also true if we replaced the Smith selection rule with any other selection rule of coflows within set $\F_k$ as well. 

We now show that the output of the algorithm belongs to the set of solutions of \ref{eq:primal}. 
Furthermore, the approximation properties of the proposed algorithmic solution rely on the dual formulation 
\begin{align}
\mbox{maximize:} \; & \sum_{\ell \in \Ll} \sum_{A \subseteq \C} y_{\ell,A}  f_{\ell}(A)  -  \sum_j \alpha_j D_j \label{eq:dual}\tag{Dual LP}\\
	 		\mbox{subj. to:} \;& \sum_{\ell \in \Ll} \sum_{A \subseteq \C : j \in S}  y_{\ell,A} \, p_{\ell,j} = w_j + \alpha_j, \quad j \in \C \label{eq:constr}\\
			\;&  y_{\ell,A}\geq 0, \;  \alpha_j\geq 0
\end{align}
If \ref{eq:primal} is feasible, the output of \cofair provides one solution for \ref{eq:primal} and one solution for \ref{eq:dual}. 

Finally, the following result demonstrates that the primal-dual iteration \cite{mastrolilli} can be extended to account for primal feasibility, i.e., the output of \cofair is a primal-feasible $\sigma$-order scheduler if one exists. Otherwise, the algorithm provides a test of feasibility for \ref{eq:primal}.
\begin{thm}\label{thm:feasible}
Let \ref{eq:primal} be feasible. \cofair produces a primal-feasible solution $\{C_k\}$ for \ref{eq:primal} and a feasible solution $\{y_{\ell,A}\}$ for \ref{eq:dual}. Let \ref{eq:primal} be not feasible, 
\cofair returns $\emptyset$. 
\end{thm}

\subsection{Output completeness}
Under feasibility conditions, the algorithm can always produce a solution corresponding to a dual solution for $\alpha =0$. Nevertheless, it is possible to use the multipliers of the dual problem in order to obtain any target primal-feasible $\sigma$-order.
\begin{thm}\label{thm:completeness}
For every primal-feasible $\sigma$-order $\sigma^*$, there exist multipliers  $\{\alpha_j\}$,  $\max \alpha_j = 1$ so that $\sigma^*$ is the output of Alg.~\ref{alg:greedy} under rescaled weights $\{\kappa \cdot w_j\}$, for some $0\leq \kappa\leq 1$.
\end{thm}

\subsection{Approximation factor}
Let $C_j$ be the output of Alg.~\ref{alg:greedy}. Let define $\Copt{j}$, $j\in \C$ a solution of \ref{eq:dynsched} and $\Clp{j}$ an solution of \ref{eq:primal}. We also define $\Csig{j}$ the coflow completion times of a $\sigma$-order using a priority output of Alg.~\ref{alg:greedy}. $\Copt{j}$ is defined  accordingly. Let recall a few results: 
\begin{lma}\label{lma:results}
i. $\sum_{k=1}^{N} w_k  \Clp{k} \leq  \sum_{k=1}^{N} w_k \Copt{k}$\\
ii. $\Big ( \sum_{j=1}^k  p_{\mu_k\sigma(j)} \Big )^2  \leq f_{\ell}(S)$, for $\ell \in \Ll$ and $S\subseteq \C$ {\rm \cite{mastrolilli} Lemma 3.2}\\
iii. $\Csig{j} - r_j \leq 2 (C_j - r_j)$ {\rm \cite{agarwal2018sincronia} Lemma 3}
\end{lma}
\begin{thm}\label{lma:primal}
Alg.~\ref{alg:greedy} produces a primal-feasible schedule such that for the corresponding $\sigma$-order coflow schedule $\sigma^*$  
\begin{equation}
\sum w_j \Ssig{j} \leq  4 \sum w_j \Sopt{j} + 4 \sum \alpha_j  D_j
\end{equation}
Assume $\alpha_j=0$ and $\sigma^*$ is feasible for \ref{eq:dynsched}, then it is a $4$ approximation w.r.t. an optimal scheduler.
\end{thm}

\subsection{Smith rule is Gaussian elimination}

The proof of Thm.~\ref{thm:feasible} described in the Appendix offers a simple interpretation from basic linear algebra of the fact that the Smith rule is always leading to a near-optimal, primal-feasible $\sigma$-order. Ultimately, the iterative selection of a pivot bottleneck-coflow pair and the weight update operated according to the Smith rule can be seen as simultaneous Gaussian elimination and selection of a $N\times N$ sub-matrix over a suitable rectangular matrix of size $N \times M\, 2^N$ corresponding to the constraints of the dual problem. In turn the existence of a coflow to be selected at each step is granted by the feasibility of \ref{eq:primal}, e.g., when using the output of \mps as input slowdown parameter $\E$. This permits to extend the proof of the result in \cite{mastrolilli}[Alg. 3.1] because under feasible slowdown constraints there is no need to update the weights of non tail-feasible coflows at each step. Furthermore, if one gives up the selection of the most charged bottleneck in the {pivot\_bottleneck} selection and Smith rule, the output is still primal-feasible, but Thm.~\ref{lma:primal} may not hold. 


\section{Numerical results}\label{sec:numerical}


This section reports on a set of extensive numerical experiments validating the proposed coflow fairness framework and the related algorithms. The numerical have been performed on a \matlab coflow scheduling simulator equipped with a coflow generator\footnote{The simulator is Open Source and it will be made available on GitHub; to obtain the source code of the simulator the reader is encouraged to contact the authors of the manuscript.}. For the algorithms based on $\sigma$-order, namely Sincronia and \cofair, the rate allocation is based on the greedy procedure. It assigns full priority to coflow $k$ over coflow $h$ on all ports where they are both active if $\sigma(k)\leq \sigma(k+1)$.  
This rate allocation has the advantage of simplicity of implementation, at the price of some performance loss against, for instance, the rate allocation adopted in Utopia which permits some rate back-filling for coflows in lower priority.

{\it Sample Coflow batches.} As indicated in the analysis of Facebook traces reported in \cite{chowdhury2015coflow}, a typical coflow pattern observed in data centers may comprise a significant fraction of small coflows but also coflows with large {\it width} (i.e., number of flows). These type of instances are denoted {\it wide-narrow} coflows batches (WN). The first set of coflow batches considered has a fraction $q$ of coflows whose width is drawn uniformly from $M/3,\ldots,M$ ($M$ being the number of ports) and a fraction $1-q$ of coflows containing single flows. The second set of coflow batches are the Map-Reduce type (MR). They are classified according to the number of mappers and the number of reducers, which are drawn from a uniform distribution in $[1,m]$ and $[1,r]$, respectively. Each reducer fetches data from each mapper. For all coflow samples, individual flow volumes are exponentially distributed with average $10$ and standard deviation $3$ in normalized traffic volume units. Links are normalized with capacity $1$ traffic volume units per second. 

{\it Performance and fairness metrics.} The metrics employed for numerical validation are the CCT,  the experimental slowdown and the stretch index (SI). The experimental slowdown for a tagged coflow $j $ is $\widehat E_j=\phi_j C_j/C_j^0$: it embraces the plain slowdown for $\phi_j=1$ and the port-occupation one for $\phi=V_j$. 
Finally, for a tagged coflow $j$, the stretch index (SI), i.e., $SI:=\sum \max( 0, (\widehat E_j/ \E) -1)$, quantifies the relative magnitude of violations w.r.t. the target slowdown constraint parameter $E$. 

\subsection{Minimum slowdown estimation.} 
\begin{figure}[t]
    \centering
    \begin{minipage}{4.2cm}
	\includegraphics[scale=0.24]{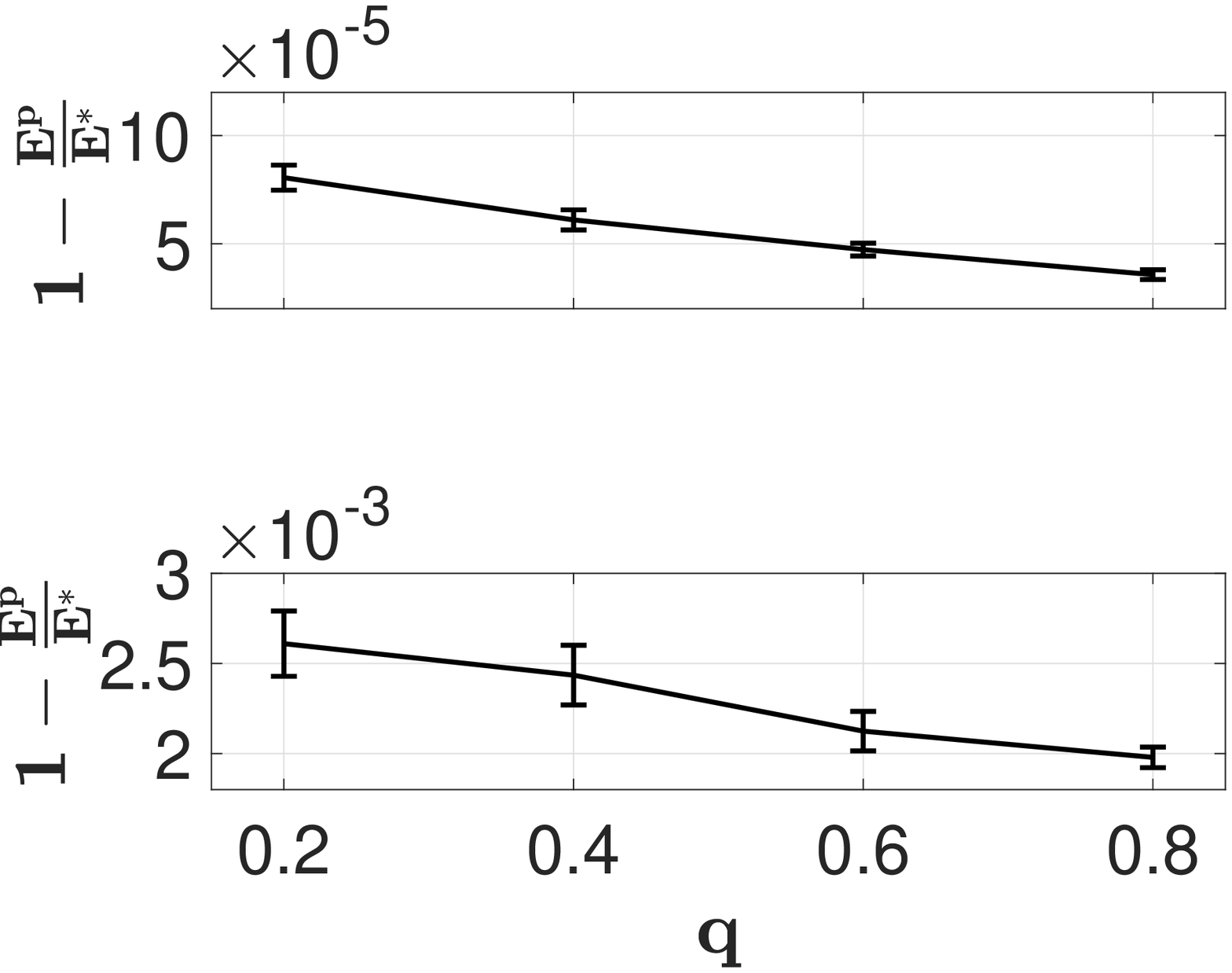} 
	\put(-30,32){\scriptsize $\phi_j=1$} \put(-30,80){\scriptsize $\phi_j=V_j$} 
	\end{minipage}\hskip2.2mm
    \begin{minipage}{4.2cm}
	\includegraphics[scale=0.24]{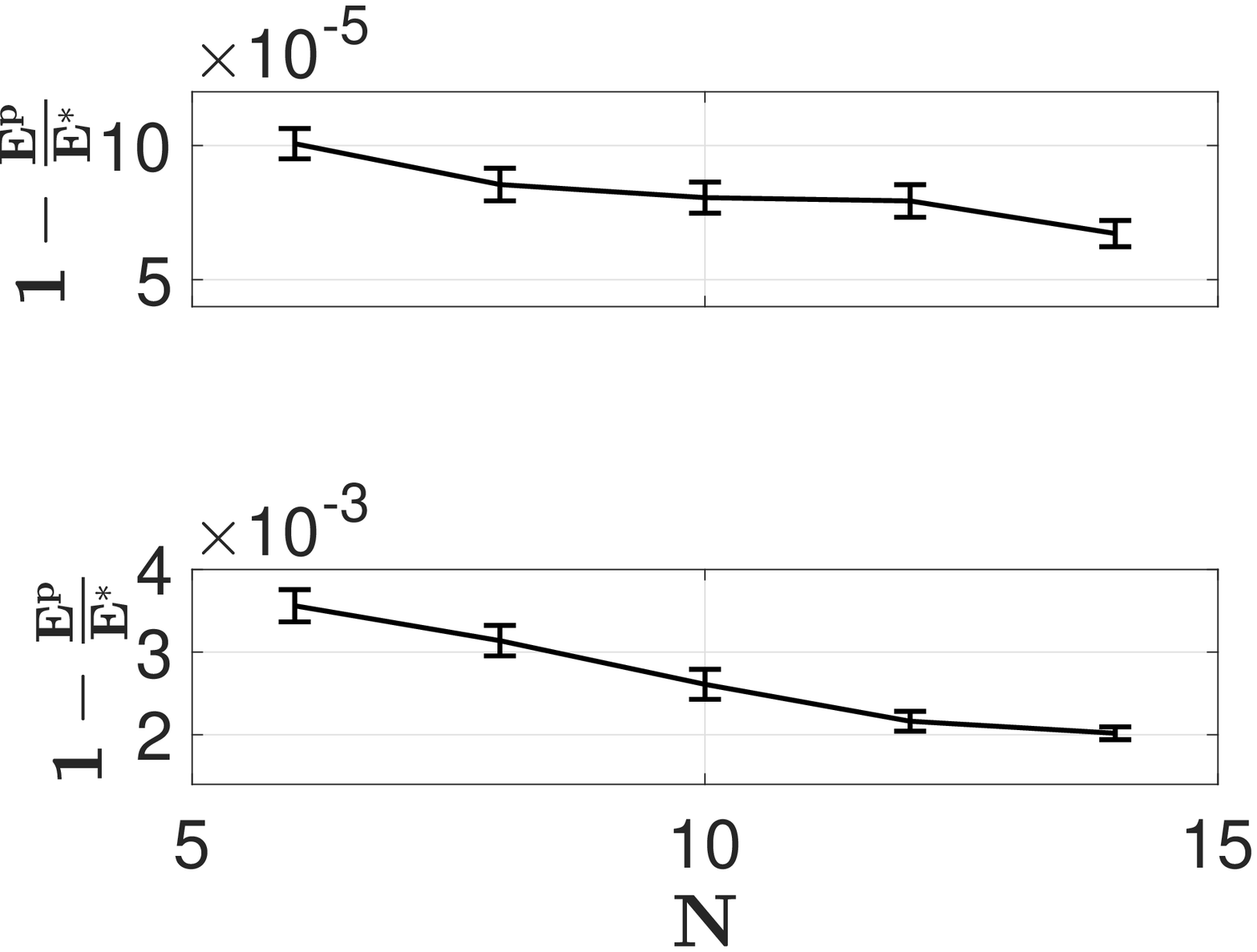} 
	\put(-31,32){\scriptsize $\phi_j=1$} \put(-31,80){\scriptsize $\phi_j=V_j$} 
	\end{minipage}
    \put(-242,-40){a)}\put(-115,-40){b)}
	\caption{Relative error on the minimum slowdown for $N=M=10$: a) increasing $q$ b) increasing $N$ and increasing $N$ for fixed $q=0.2$.}\label{fig:num1}
\end{figure}
As described in Sec. \ref{sec:slowdown}, the minimum slowdown $\E^*$ corresponds to the largest average progress that can be ensured to all the coflows in a batch. Fig. \ref{fig:num1}a and \ref{fig:num1}b reported on the relative error which is obtained by approximating $\E^*$  using the value attained for the primal value $\E^p$ generated as output of \mps. Each point of those graphs is obtained by averaging the relative error obtained over $100$ batch instances, where the LP solution corresponds to a time slot of $1$ second. 

Fig. \ref{fig:num1}a reports on the approximation results in the case or increasing values of $q$, $M=N=10$, whereas Fig. \ref{fig:num1}b reports on the same experiment for increasing values of $N$. In both cases, the approximation improves for larger values of $q$ and $N$, respectively. The important information is that the approximation improves the higher the fabric congestion, and it appears very tight, i.e., the relative error is bounded below $10^{-4}$ for $\phi=V_j$ and $3\times 10^{-3}$ for $\phi=1$. These results confirm that the minimum slowdown of the primal problem $\E^p$ appears an extremely tight approximation compared against the exact value $\E^*$ obtained by solving the corresponding LP. For each set of $100$ experiments we observed at most $1$ outlier, i.e., a batch for which the relative error falls above $1\%$. By inspection, we found that this happens for very specific cases where a coflow batch has several bottlenecks of same size, and where a primal-feasible schedule may prioritize certain coflows on bottlenecks where in turn they slow down others, thus increasing the maximum slowdown value for that batch. 

\subsection{Plain slowdown: $\phi_j=1$.} 

This set of experiments considers the plain slowdown, i.e., $\phi_j=1$. With this choice, between two coflows with same volume, it is fair if the coflow with the lowest CCT in isolation finishes first. In each experiment, same coflow instances are processed using Sincronia, \cofair and Utopia. In all experiments, \cofair receives $E^*$ as output of \mps. For the plain slowdown, the Jain index $J(\mathbf R)=\frac1{K} \frac{\sum_{j=1}^K R_j^2}{(\sum_{j=1}^K R_j)^2}$  can measure how fair is the progress distribution among coflows, i.e., the average rate $R_j$. 
\begin{figure}[t]
        \centering
        \begin{minipage}{4.2cm}
	\includegraphics[scale=0.24]{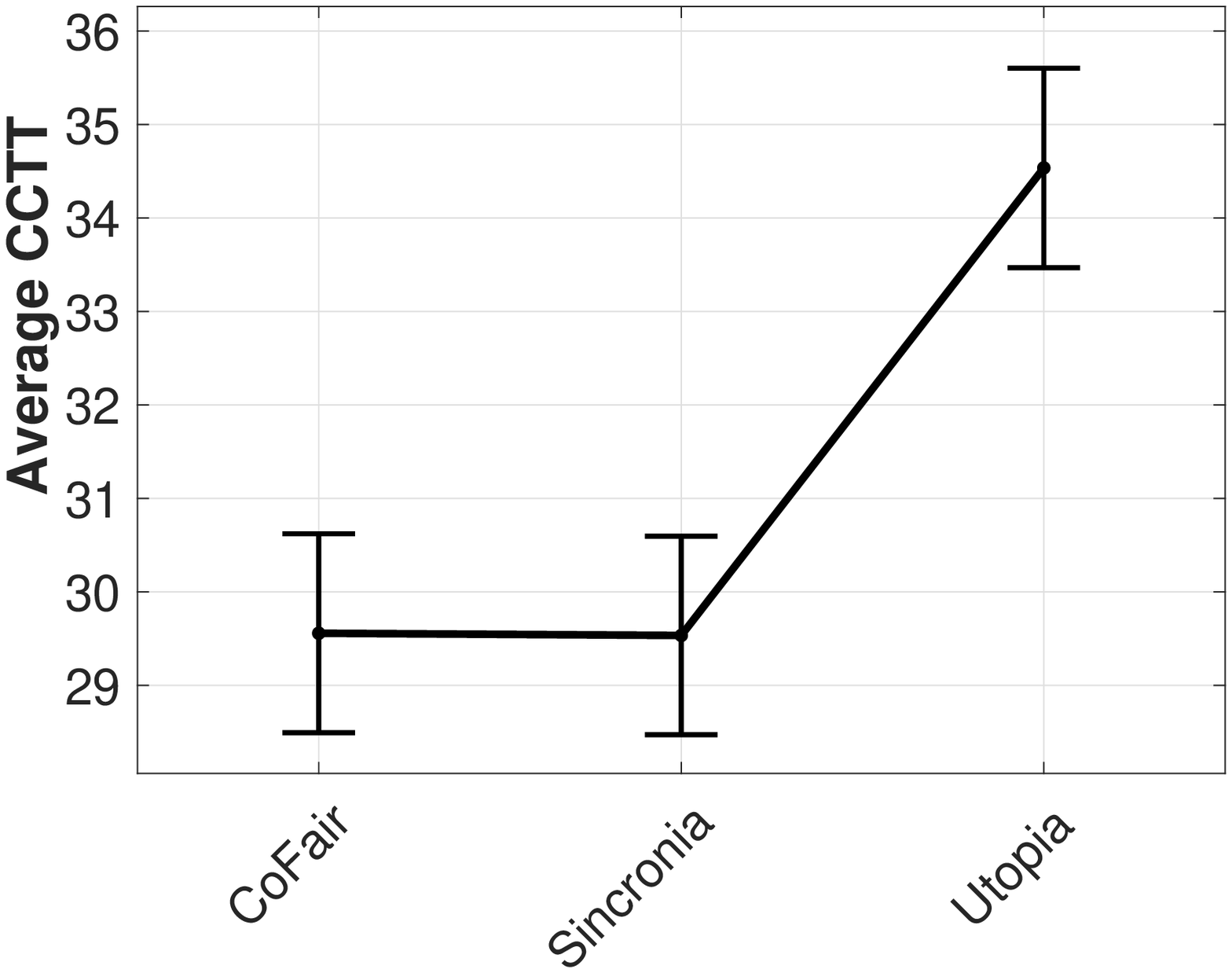} 
	\put(-123,0){a)}
	\end{minipage}\hskip1.5mm
        \begin{minipage}{4.2cm}
	\includegraphics[scale=0.24]{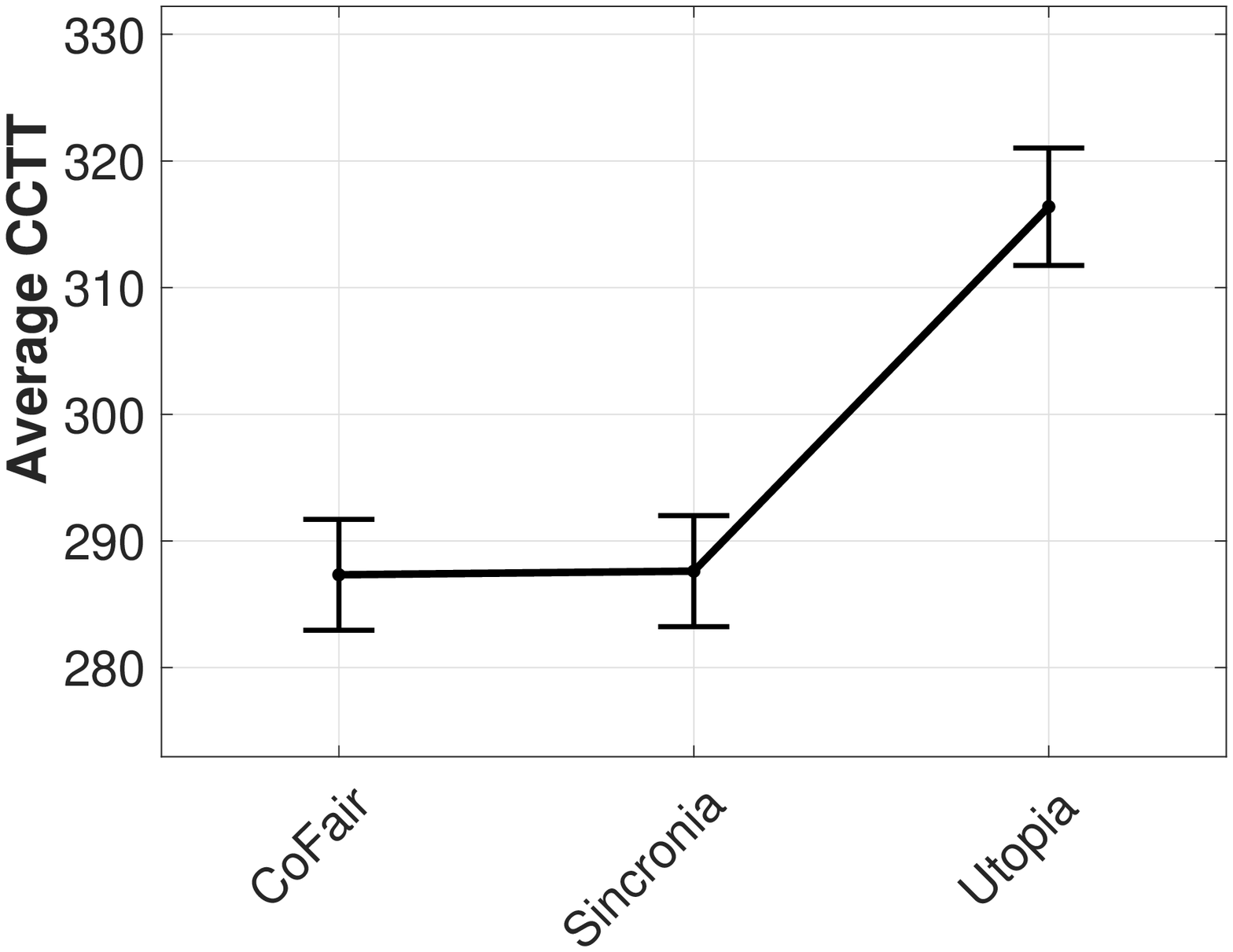} 
	\put(-125,0){b)}
	\end{minipage}\hskip2.5mm
	\caption{Average CCT normalized against Sincronia for $M=30$: a) $p=0.2$ and $N=30$ and b) $p=0.8$ and $N=100$; $95\%$ confidence intervals are superimposed.}\label{fig:num1bis}
\end{figure}

\begin{figure*}[t]
    \centering
    \begin{minipage}{4.2cm}
	\includegraphics[scale=0.24]{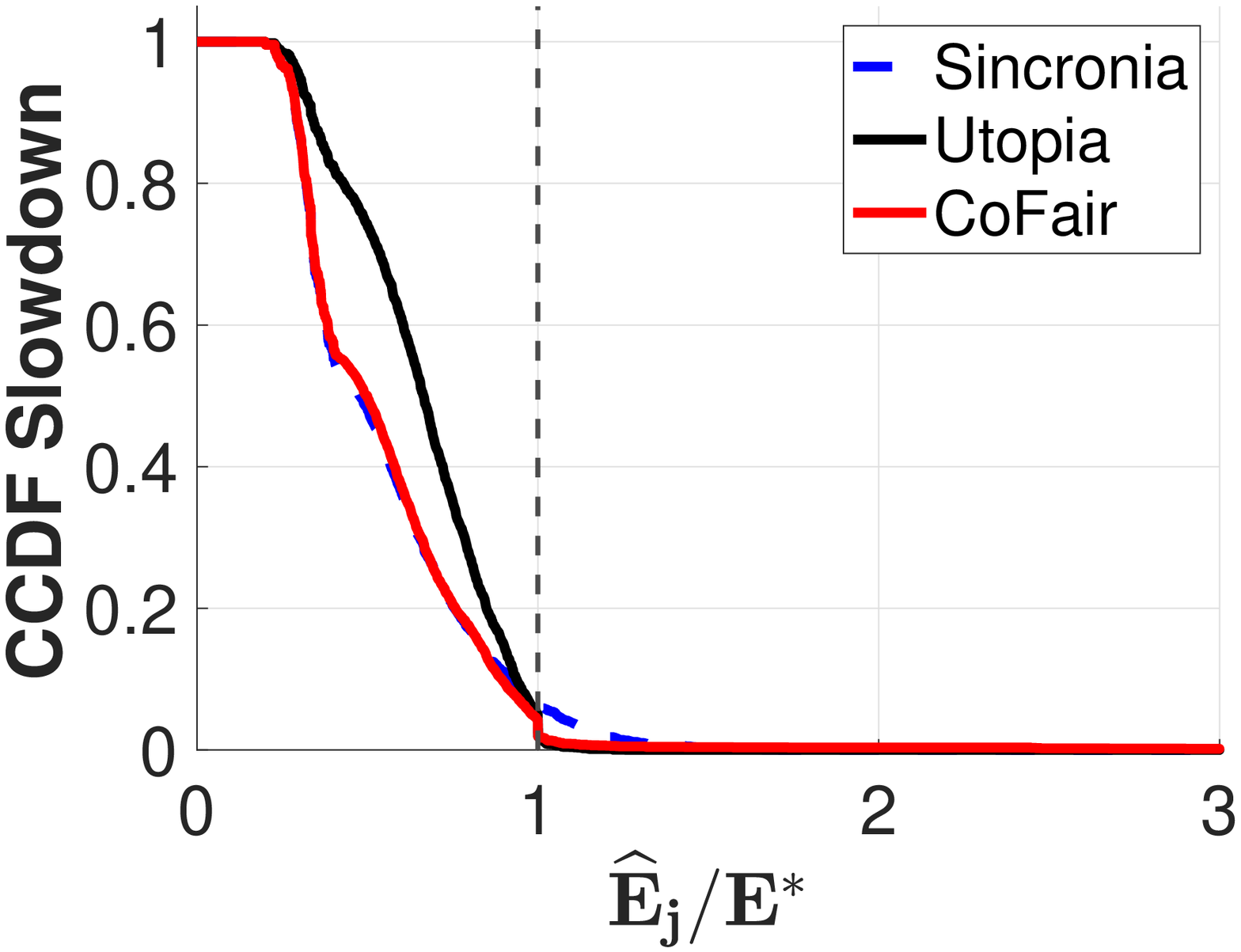} 
	\put(-121,0){a)}
	\end{minipage}\hfill
	\begin{minipage}{4.2cm}
	\includegraphics[scale=0.24]{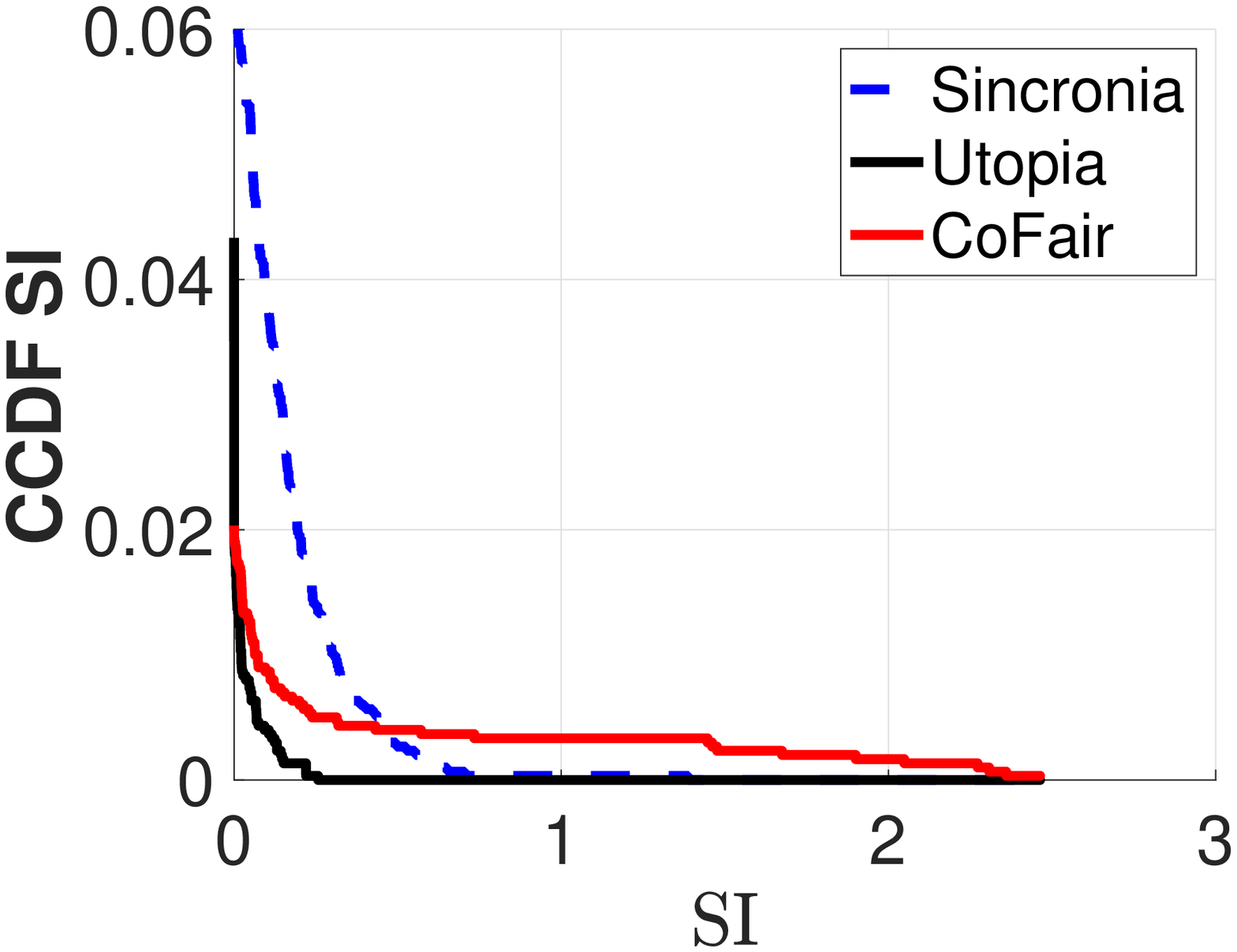} 
	\put(-121,0){b)}
	\end{minipage}\hfill
    \begin{minipage}{4.2cm}
	\includegraphics[scale=0.24]{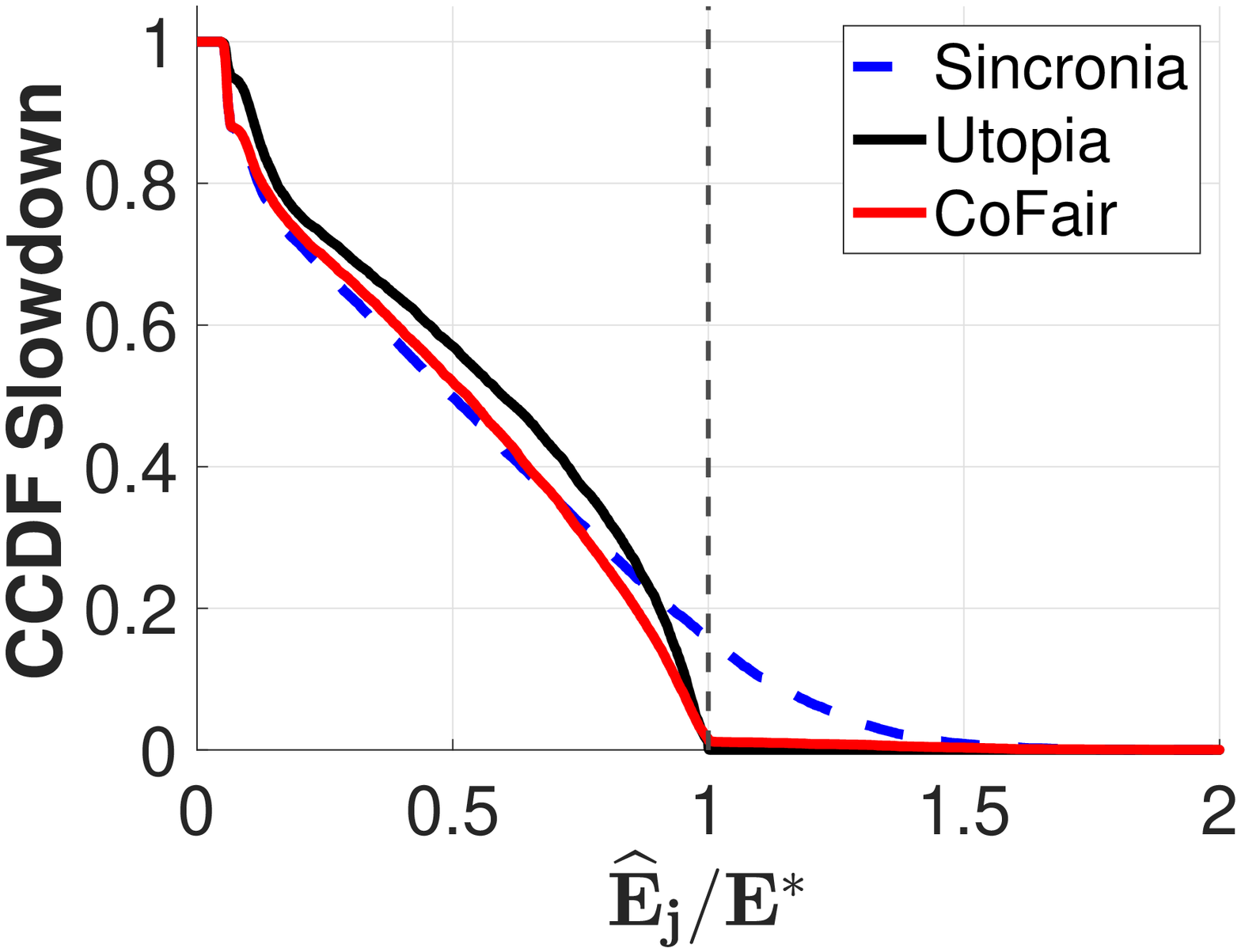} 
	\put(-121,0){c) }
	\end{minipage}\hfill
	\begin{minipage}{4.2cm}
	\includegraphics[scale=0.24]{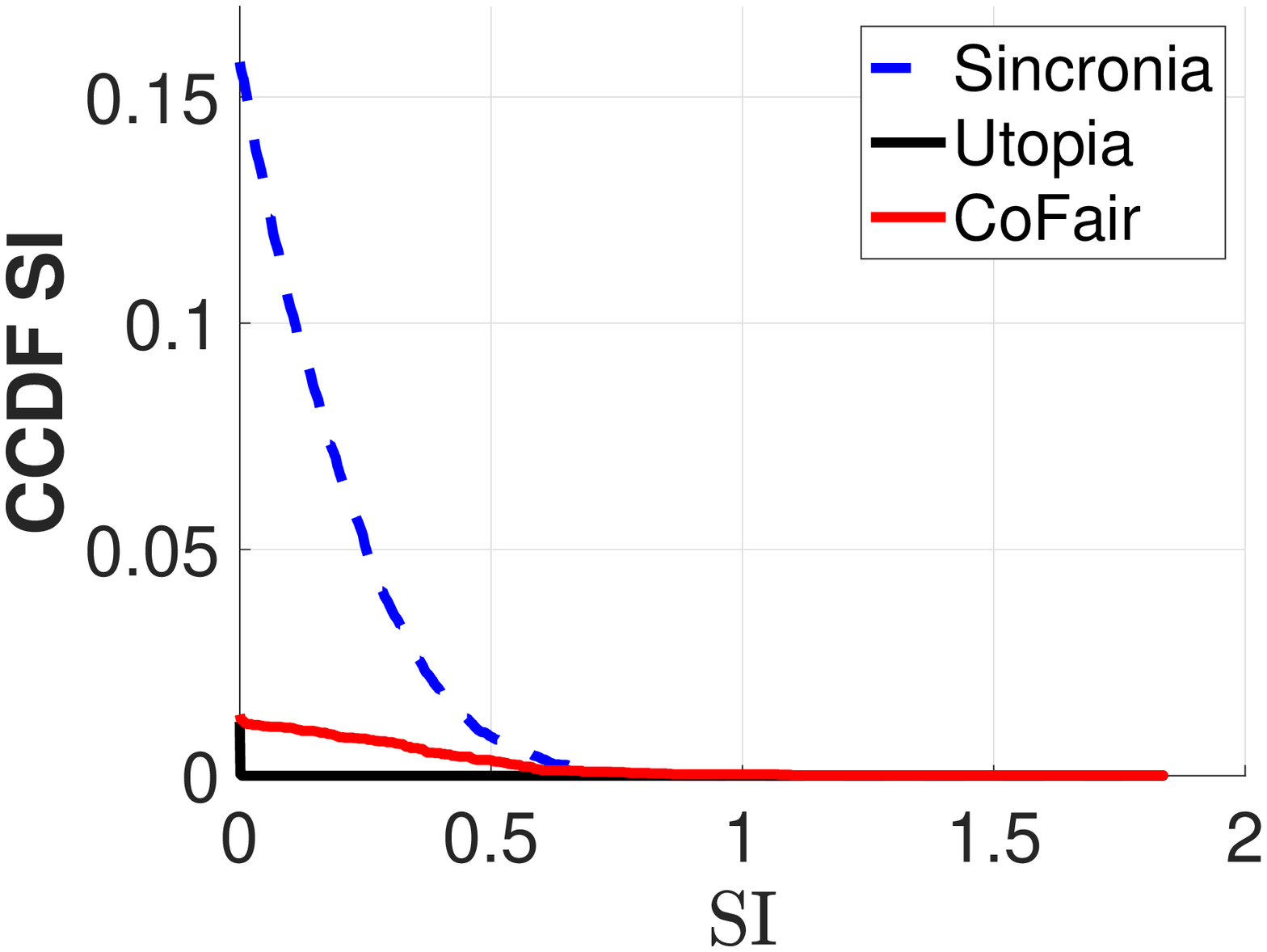} 
	\put(-121,0){d)}
	\end{minipage}\hfill
	\caption{WN traces: a) and c)  CCDF of the normalized slowdown; b) and d) CCDF of the stretch index; setting: $M=30$, $q=0.2$ and $N=30$ for a) and b); $q=0.8$ and $N=100$ for c) and d). } \label{fig:num2}
\end{figure*}

Fig.~\ref{fig:num1bis}a, Fig.~\ref{fig:num2}a and Fig. \ref{fig:num2}b refer to the performance figures measured for an experiment with $100$ batches of coflows of NW type for $p=0.2$ and $N=30$. Performance figures reported in Fig.~\ref{fig:num1bis}b and Fig. \ref{fig:num2}c and d refer to $p=0.8$ and $N=100$.
 
Fig.~\ref{fig:num1bis} reports on the average CCT attained by Sincronia, \cofair and Utopia: to this respect, the average CCT of \cofair and Sincronia coincide. On the other hand, the performance loss degradation of Utopia w.r.t. Sincronia and \cofair is significant, i.e., on the order of $17\%$ for $p=0.2$ and $9\%$ for $p=0.8$. For $p=0.2$ the Jain index of \cofair is $0.60$, for Sincronia is $0.59$ whereas Utopia only scores $0.53$. Conversely, for $p=0.8$ the Jain index of \cofair is $0.61$, for Sincronia is $0.60$ whereas Utopia scores $0.58$. 

Fig.~\ref{fig:num2} illustrates the complementary cumulative distribution function (CCDF) of the slowdown and of the stretch index. The dashed vertical line seen in Fig.~\ref{fig:num2}a and Fig.~\ref{fig:num2}c separates the {\it inner interval} on the left where $\widehat E_j \leq E^*$, i.e., no violation occurs, and the {\it outer interval} $\widehat E_j > E^*$, i.e., where the slowdown does not meet the target. As observed in Fig.~\ref{fig:num2}a, Sincronia and \cofair provide the best performance for $\widehat E_j \leq E^*$. However, as expected, it suffers a larger number of violations compared to Utopia. Finally, \cofair matches the behavior of Sincronia for $\widehat E_j\leq E^*$ and Utopia for $\widehat E_j \geq E^*$, even though the rare violations, when they occur, appear relatively larger. Same behaviour is confirmed for $p=0.8$ in Fig.~\ref{fig:num1bis}a, Fig.~\ref{fig:num2}a and Fig. \ref{fig:num2}b. In this case, due to the increased congestion, the violations of Sincronia are larger than in the previous experiments, and yet \cofair balances between a near optimal CCT minimization and slowdown guarantees. 
\begin{figure*}[t]
        \centering
         \begin{minipage}{4.2cm}
	\includegraphics[scale=0.24]{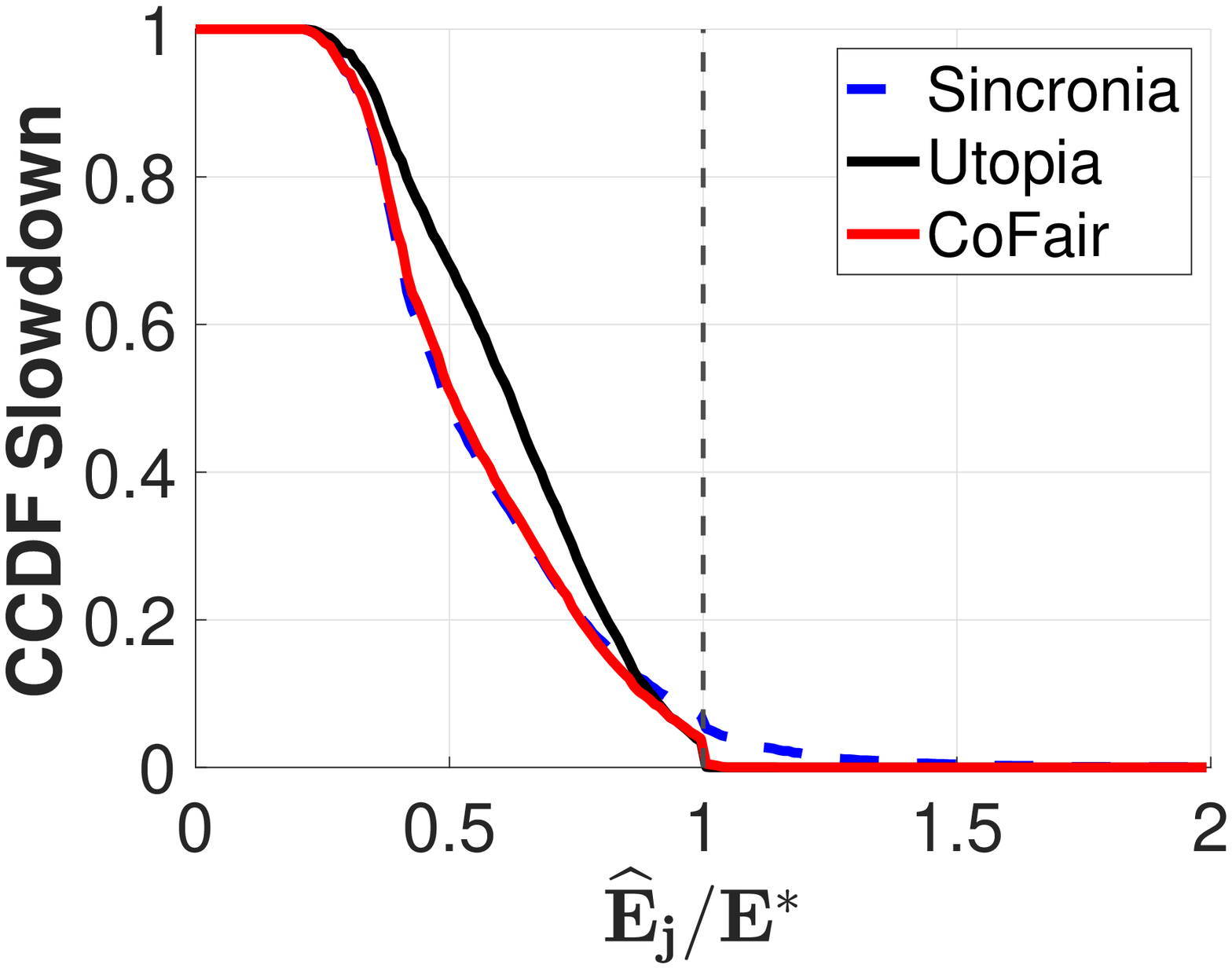} 
	\put(-125,0){a)}
	\end{minipage}
	\begin{minipage}{4.2cm}
	\includegraphics[scale=0.24]{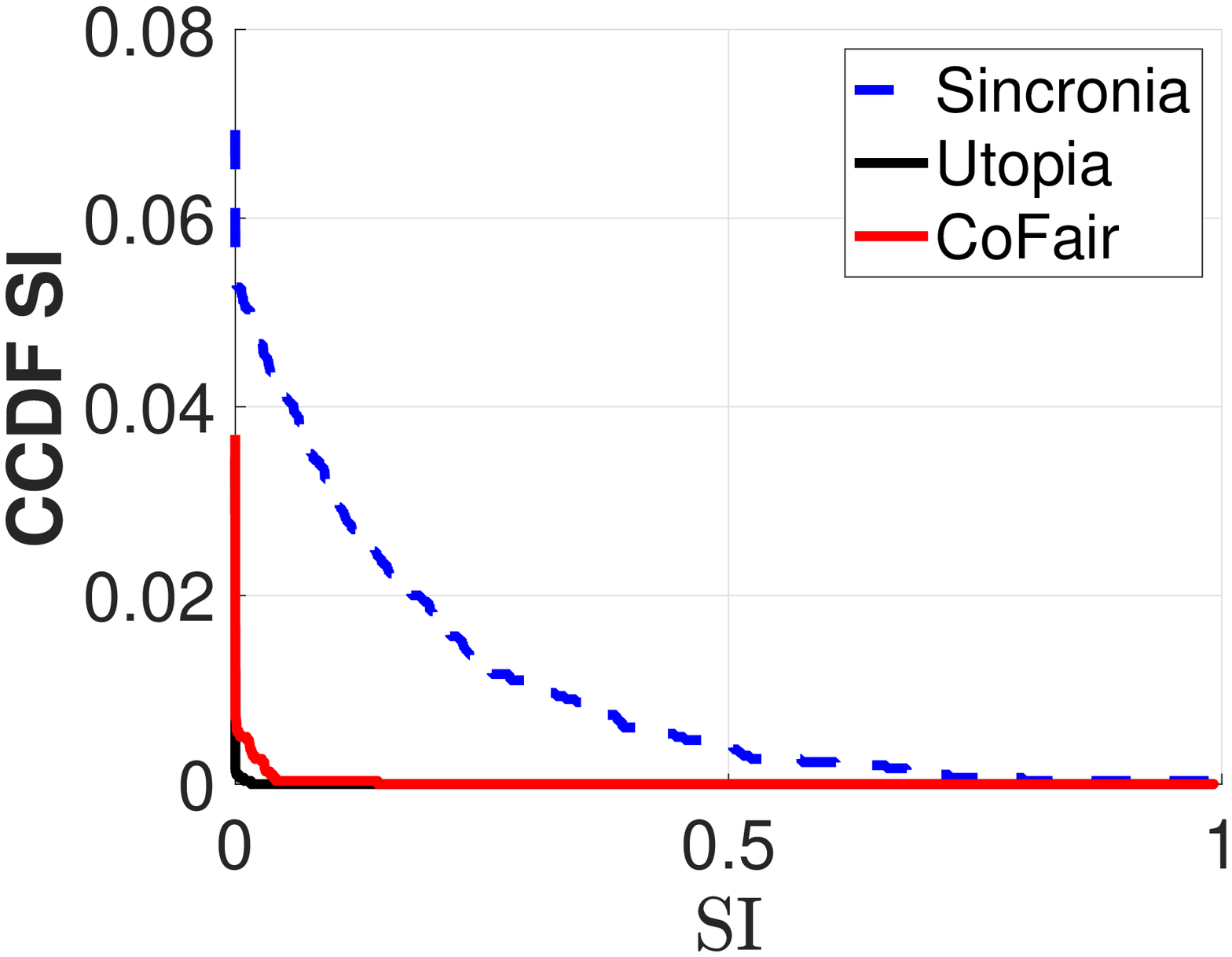} 
	\put(-125,0){b)}
	\end{minipage}
         \begin{minipage}{4.2cm}
	\includegraphics[scale=0.24]{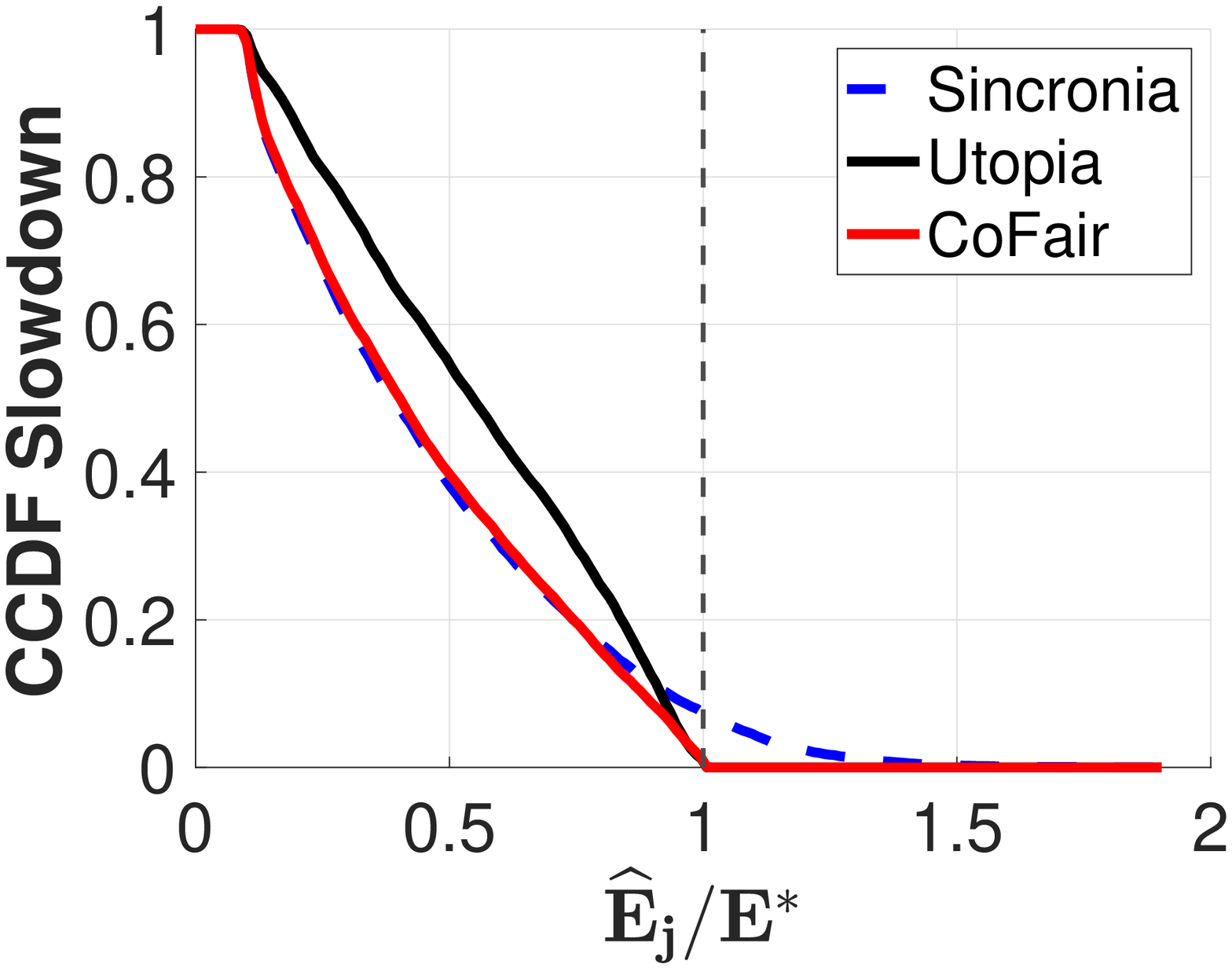} 
	\put(-125,0){c)}
	\end{minipage}
	\begin{minipage}{4.2cm}
	\includegraphics[scale=0.24]{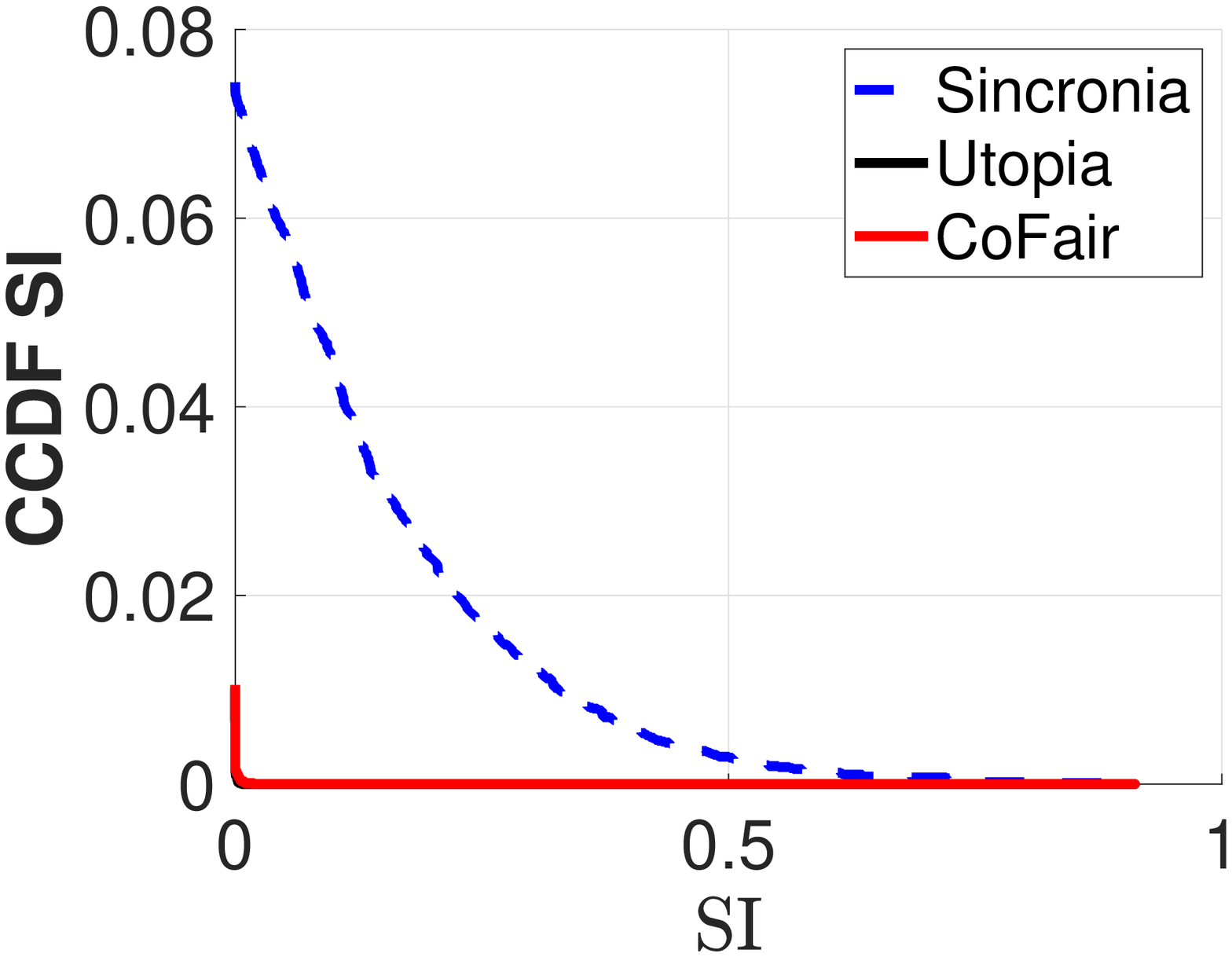} 
	\put(-125,0){d)}
	\end{minipage}
	\caption{MR traces: a) and c)  CCDF of the normalized slowdown; b) and d) CCDF of the stretch index; setting: $M=30$, and $N=30$, $m=10$ and $r=3$ for a) and b);$M=30$, and $N=100$, $m=10$ and $r=10$  $q=0.8$ and $N=100$ for c) and d).} \label{fig:num3}
\end{figure*}

The same experiments are repeated for MR coflow batches under different configurations on the number of mappers and reducers. These coflow batches generate severe congestion both at the input and at the output ports of the fabric. Fig.~\ref{fig:num3}a and Fig.~\ref{fig:num3}b refer to the case of $100$ batches of coflows with $N=30$ MR coflows where $m=10$ and $r=3$. As before, Sincronia and \cofair provide the best performance in the inner interval. However, Sincronia has a significant number of deviations, i.e., on the order of $10\%$ for coflows, whereas Utopia provides a better performance in terms of deviation. This is confirmed also by the Jain index, where \cofair scores $0.89$, Sincronia $0.88$  ad Utopia $0.87$. Ultimately, \cofair matches Sincronia in the inner interval and Utopia in the outer one, providing a better tradeoff. 

The experiment is hence performed for $m=10$ and $r=10$ and $N=100$ coflows. Again, the slowdown CCDF of \cofair is practically superimposed to that of Sincronia in the inner interval and incurs very few violations ($<10^{-3}$). With respect to the Jain index for the coflow progress, \cofair and Sincronia score $0.84$ whereas Utopia $0.80$.
\begin{figure}[t]
    \centering
	\begin{minipage}{4.2cm}
	\includegraphics[scale=0.24]{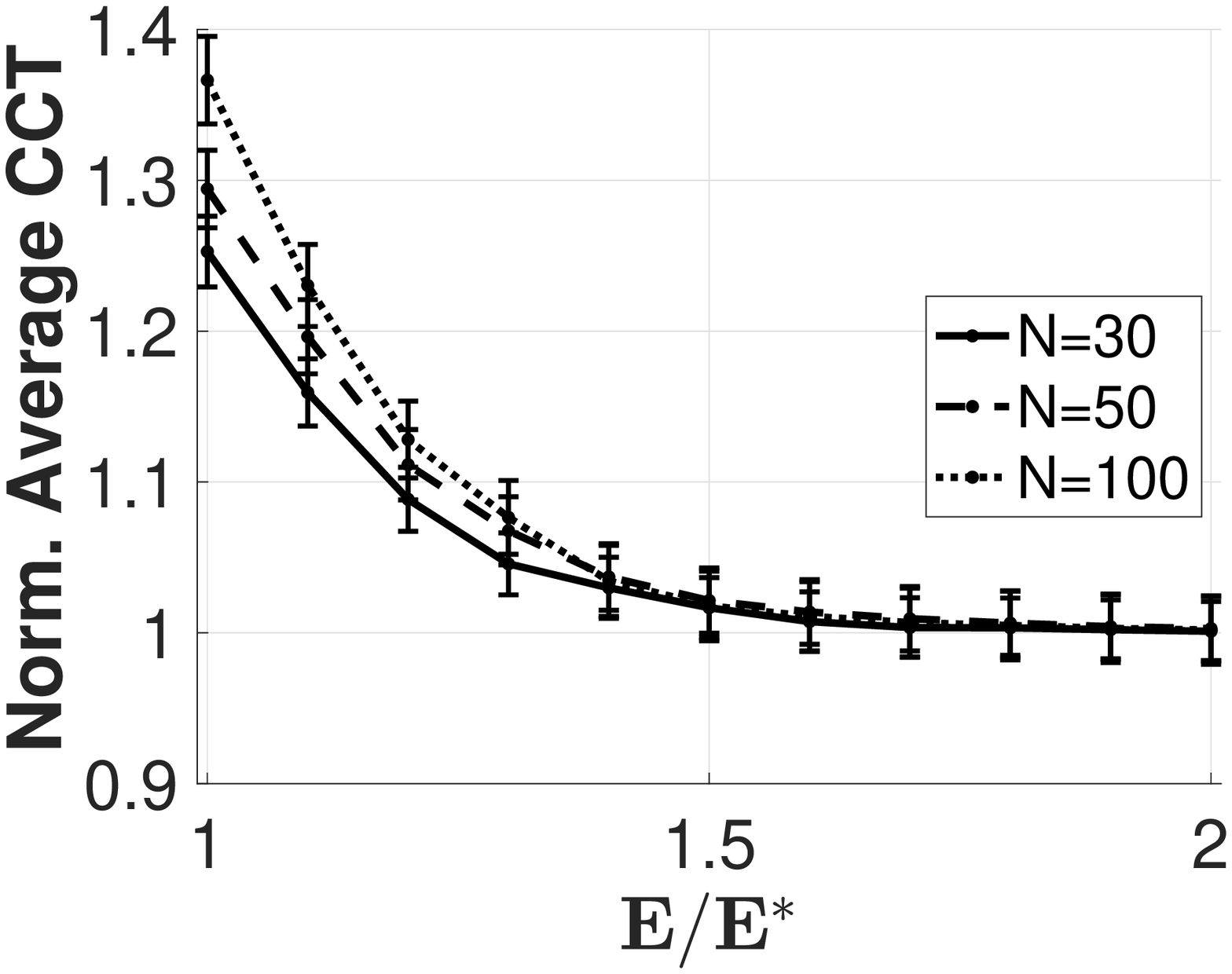} 
	\put(-123,0){a)}
	\end{minipage}\hskip1.5mm
       \begin{minipage}{4.2cm}
	\includegraphics[scale=0.25]{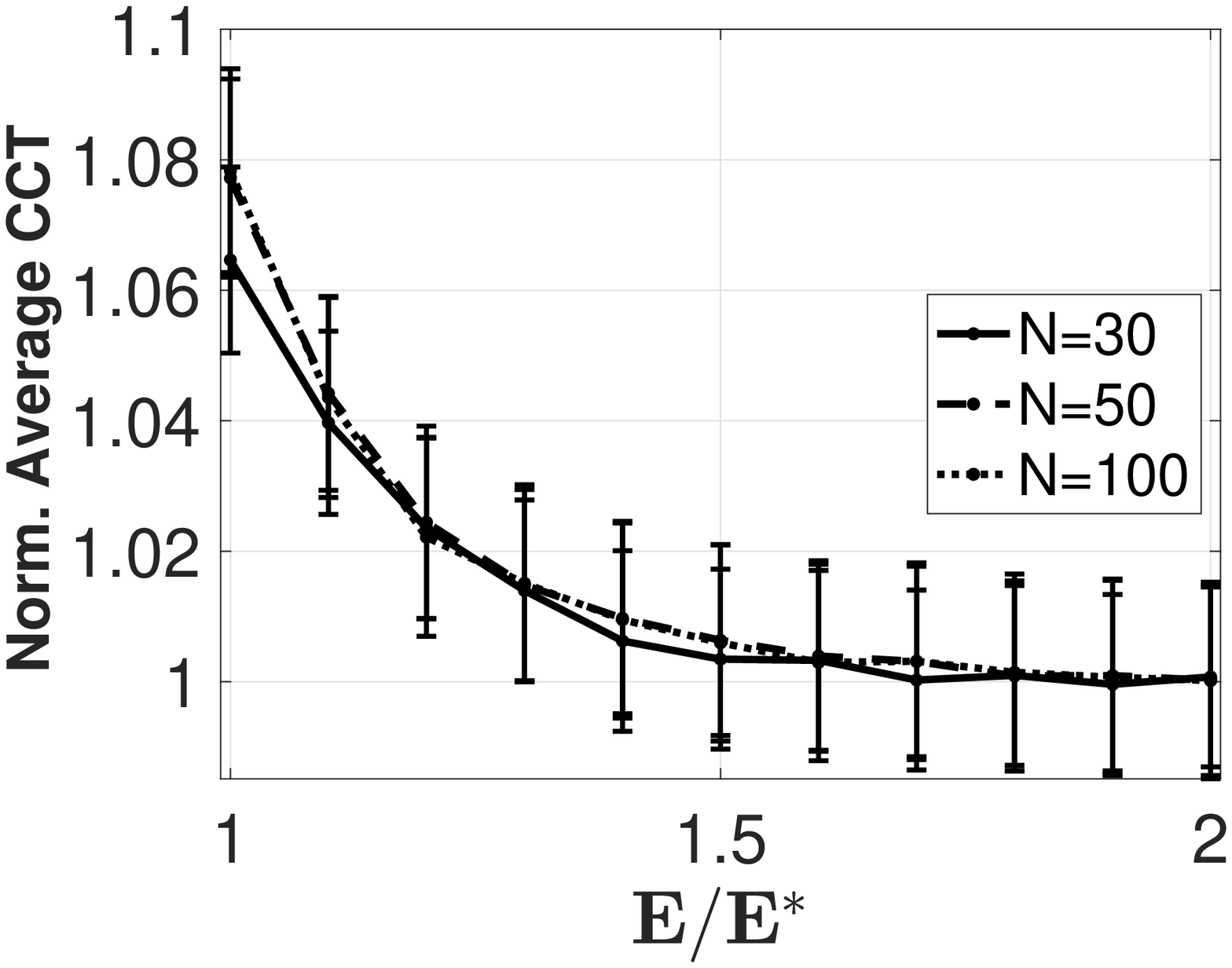} 
	\put(-125,0){b)}
	\end{minipage}
	\caption{Average CCT normalized against Sincronia for $M=30$: a) $p=0.2$ and $N=30$; b) $p=0.8$ and $N=100$.}\label{fig:num1tris}
\end{figure}

\begin{figure*}[t]
        \centering
        \begin{minipage}{4.2cm}
	\includegraphics[scale=0.24]{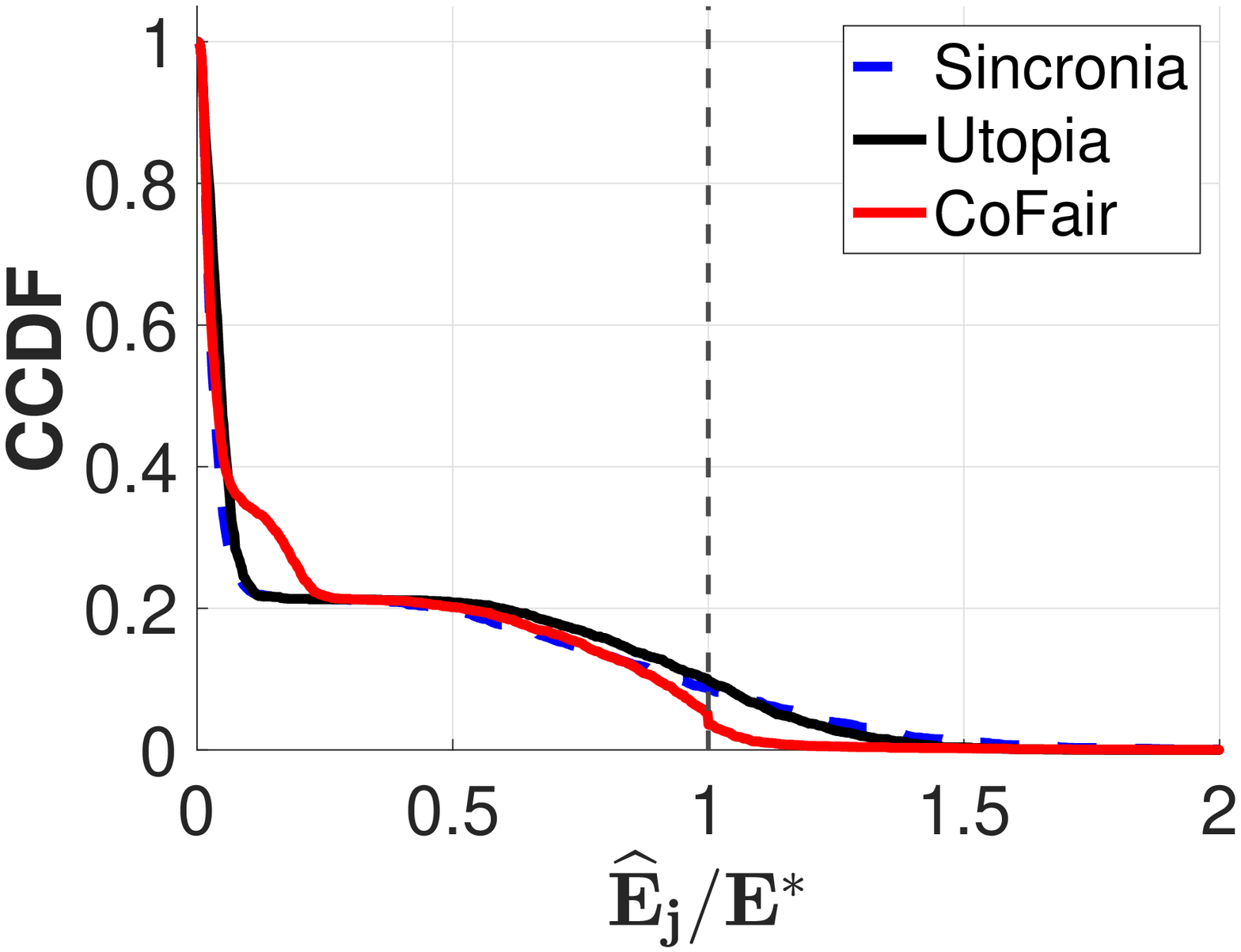} 
	\put(-125,0){a)}
	\end{minipage}
	\begin{minipage}{4.2cm}
	\includegraphics[scale=0.24]{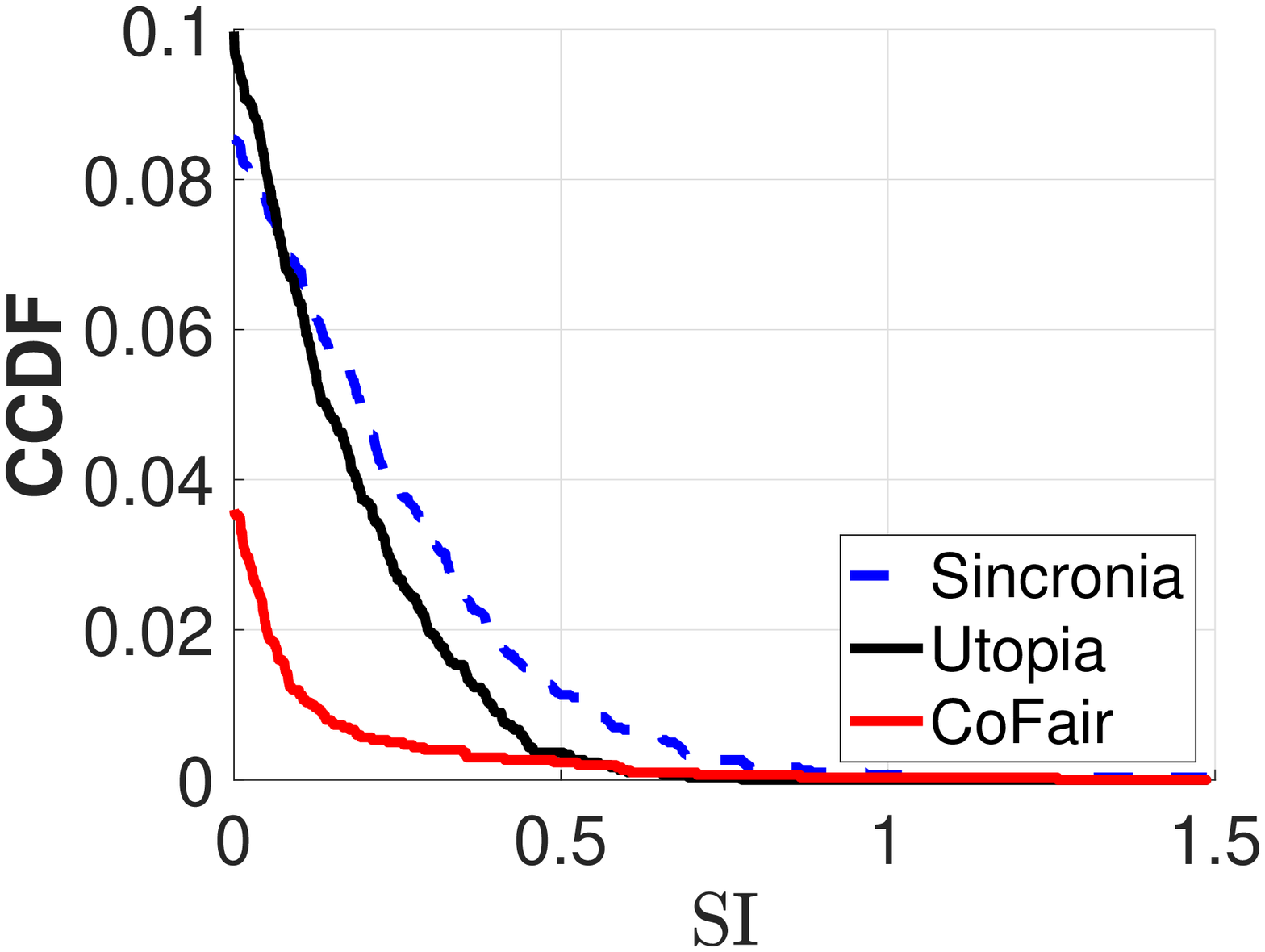} 
	\put(-125,0){b)}
	\end{minipage}
	\begin{minipage}{4.2cm}
	\includegraphics[scale=0.24]{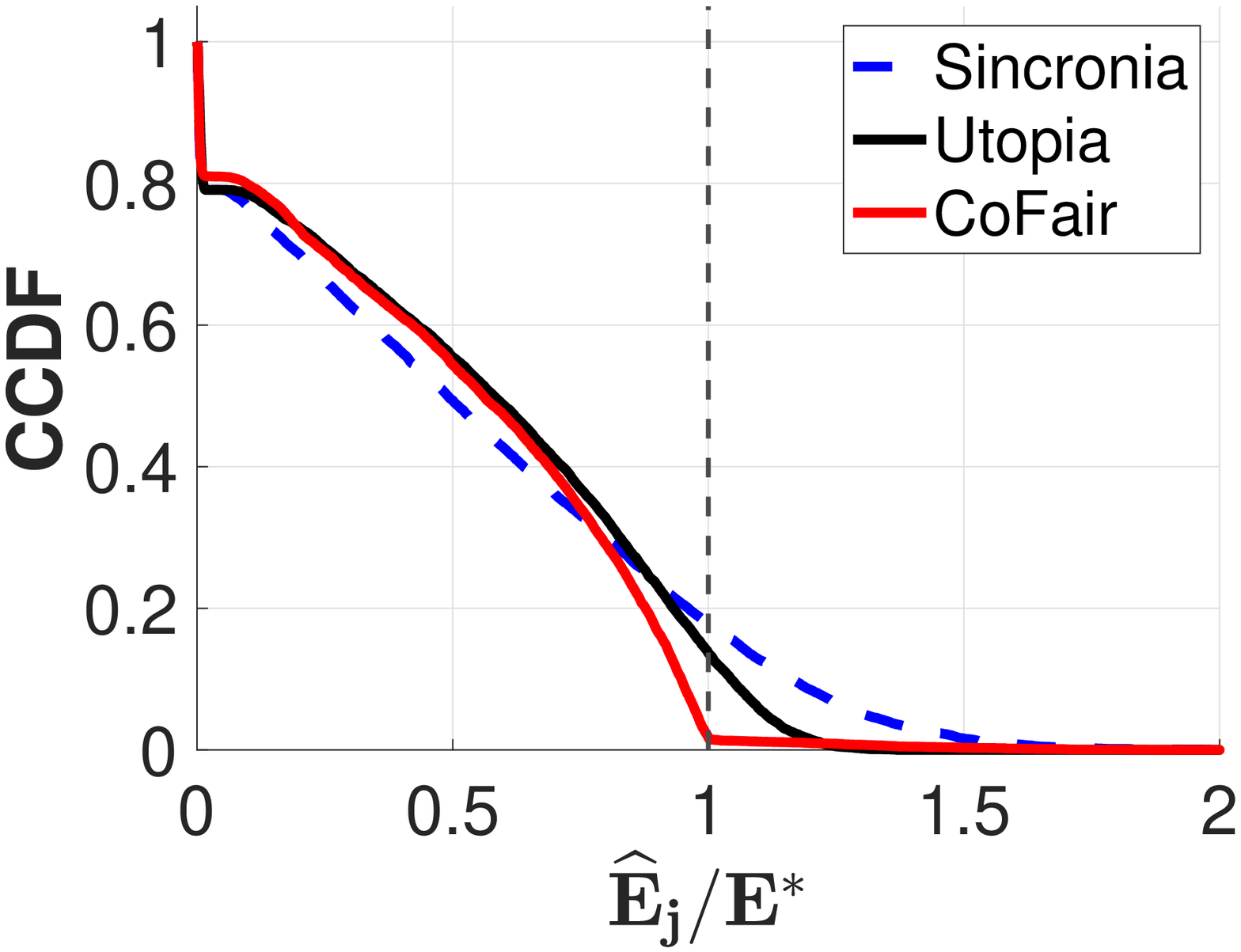} 
	\put(-125,0){c)}
	\end{minipage}
	\begin{minipage}{4.2cm}
	\includegraphics[scale=0.24]{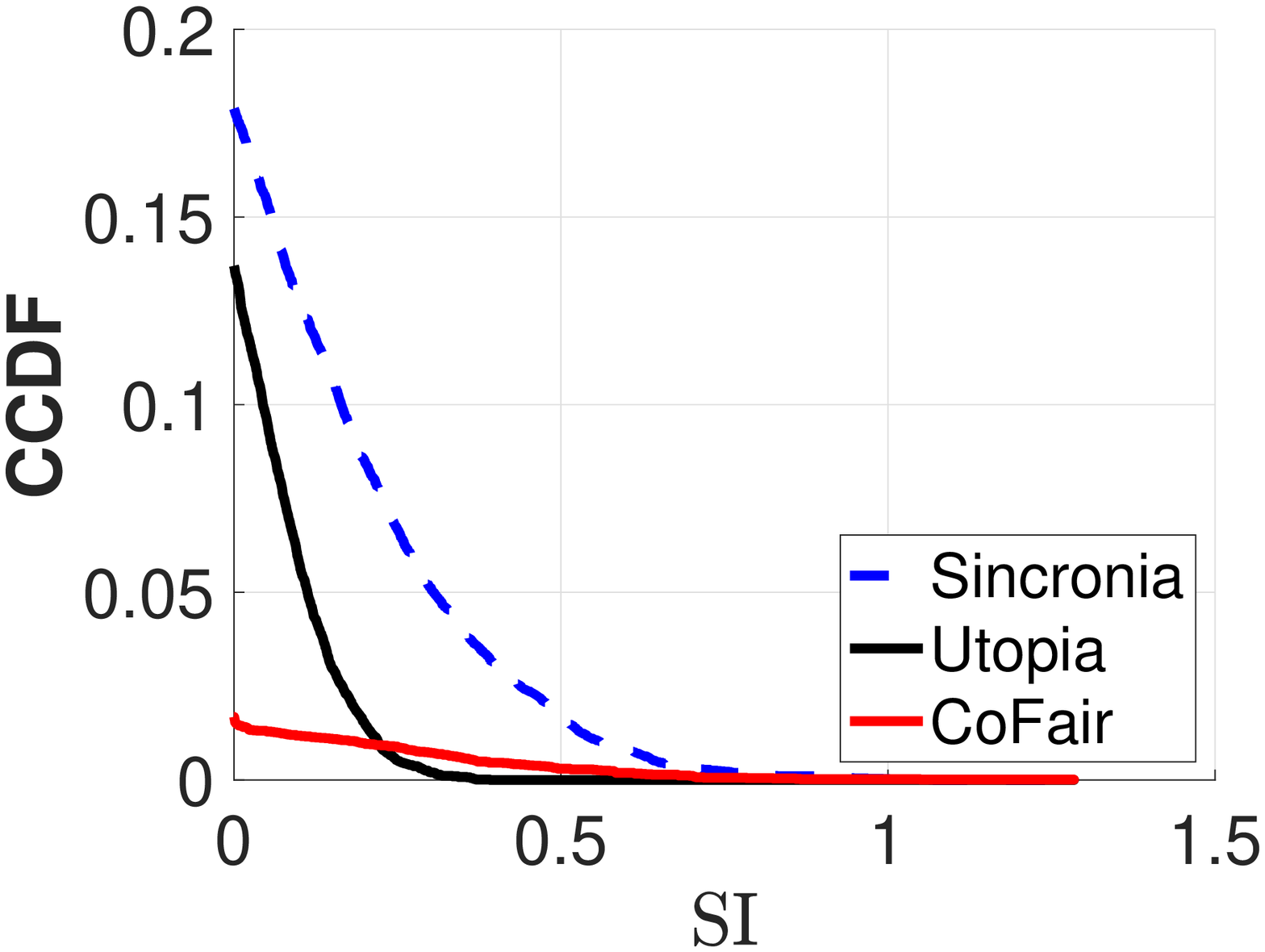} 
	\put(-125,0){d)}
	\end{minipage}
	\caption{WN traces, $\phi_j=V_j$ a) and c)  CCDF of the normalized slowdown; b) and d) CCDF of the stretch index; setting: $M=30$, $q=0.2$ and $N=30$ for a) and b); $q=0.8$ and $N=100$ for c) and d).} \label{fig:num4}
\end{figure*}

\begin{figure*}[t]
        \centering
         \begin{minipage}{4.2cm}
	\includegraphics[scale=0.24]{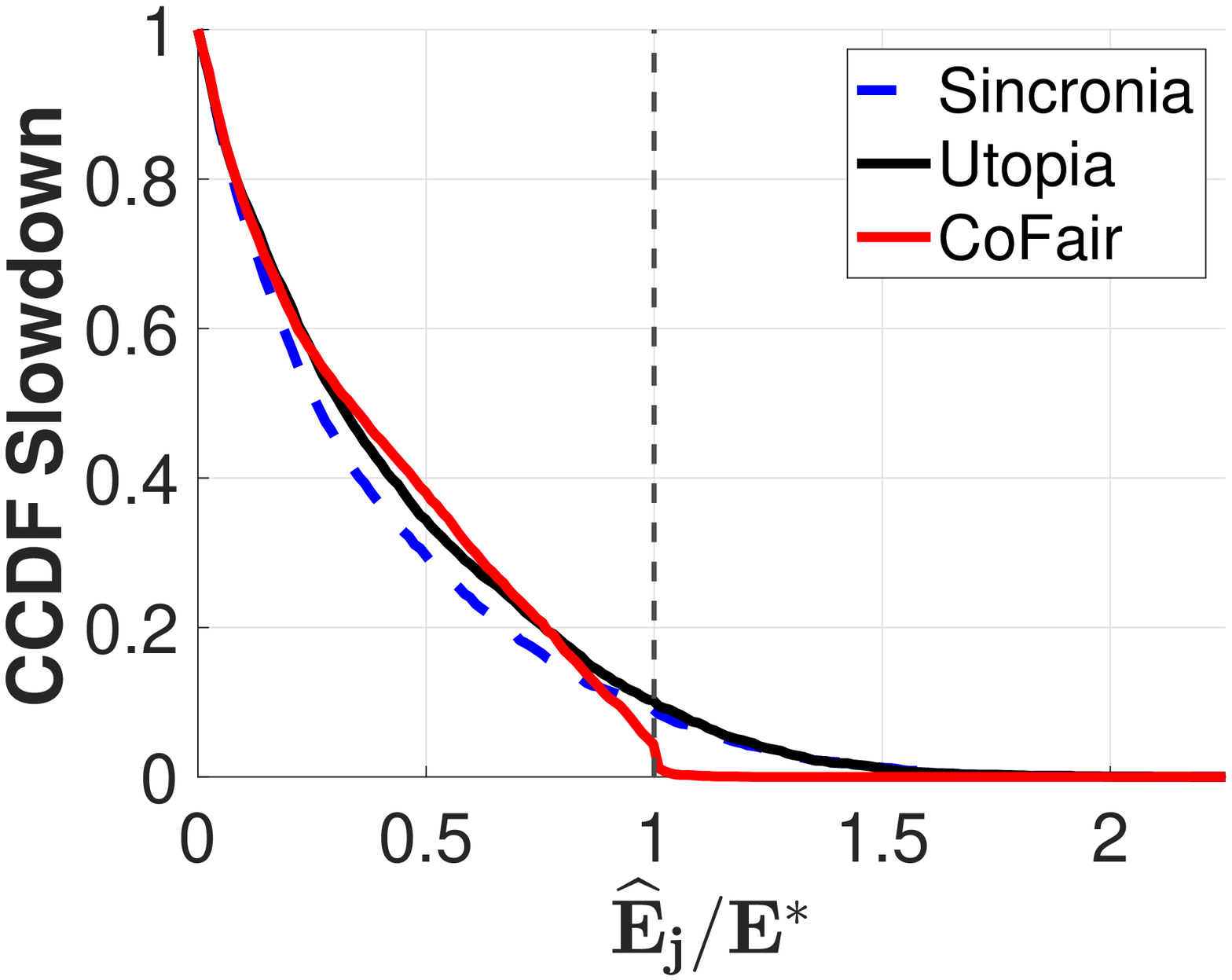} 
	\put(-125,0){a)}
	\end{minipage}
	\begin{minipage}{4.2cm}
	\includegraphics[scale=0.24]{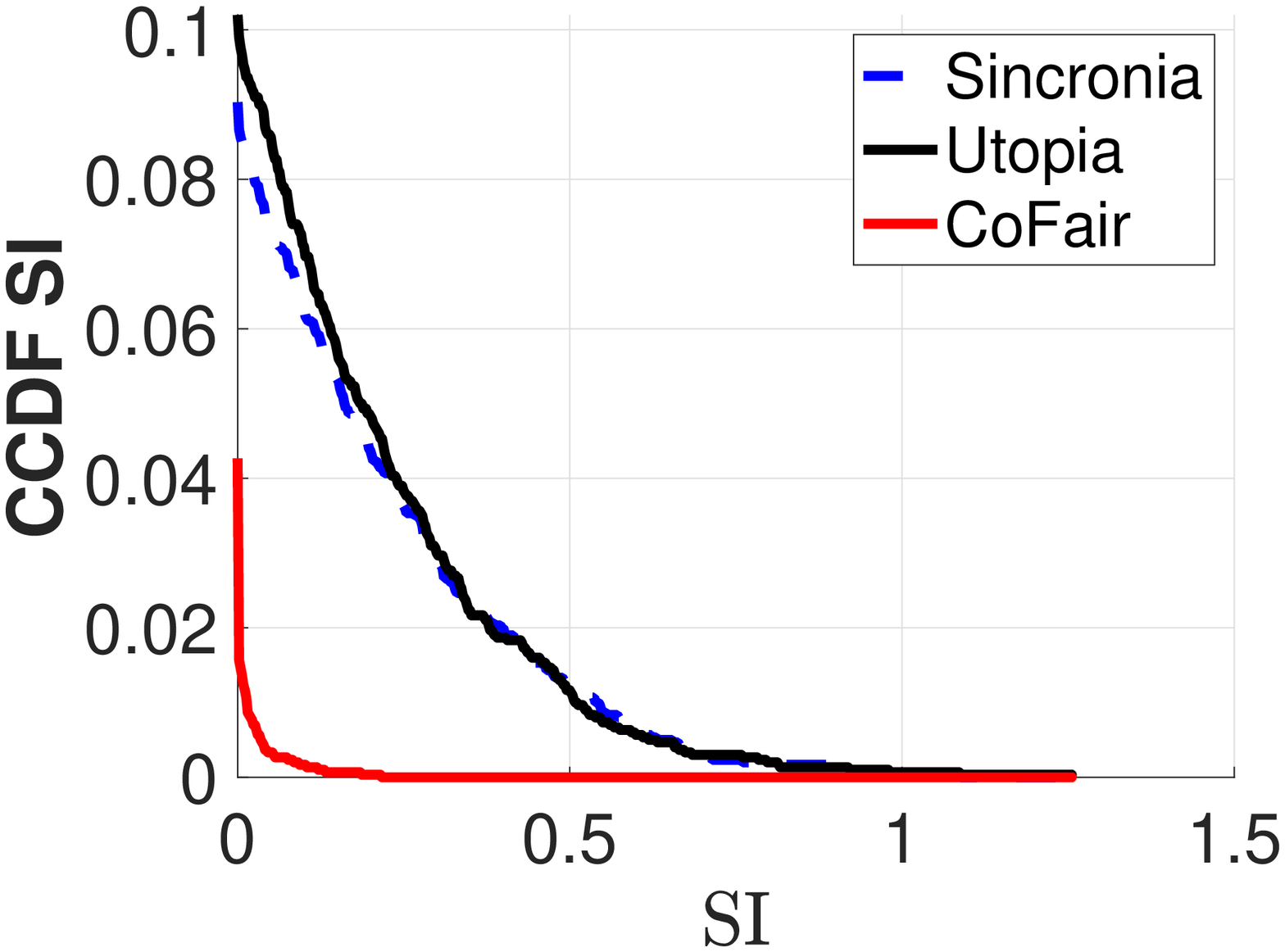} 
	\put(-125,0){b)}
	\end{minipage}
         \begin{minipage}{4.2cm}
	\includegraphics[scale=0.24]{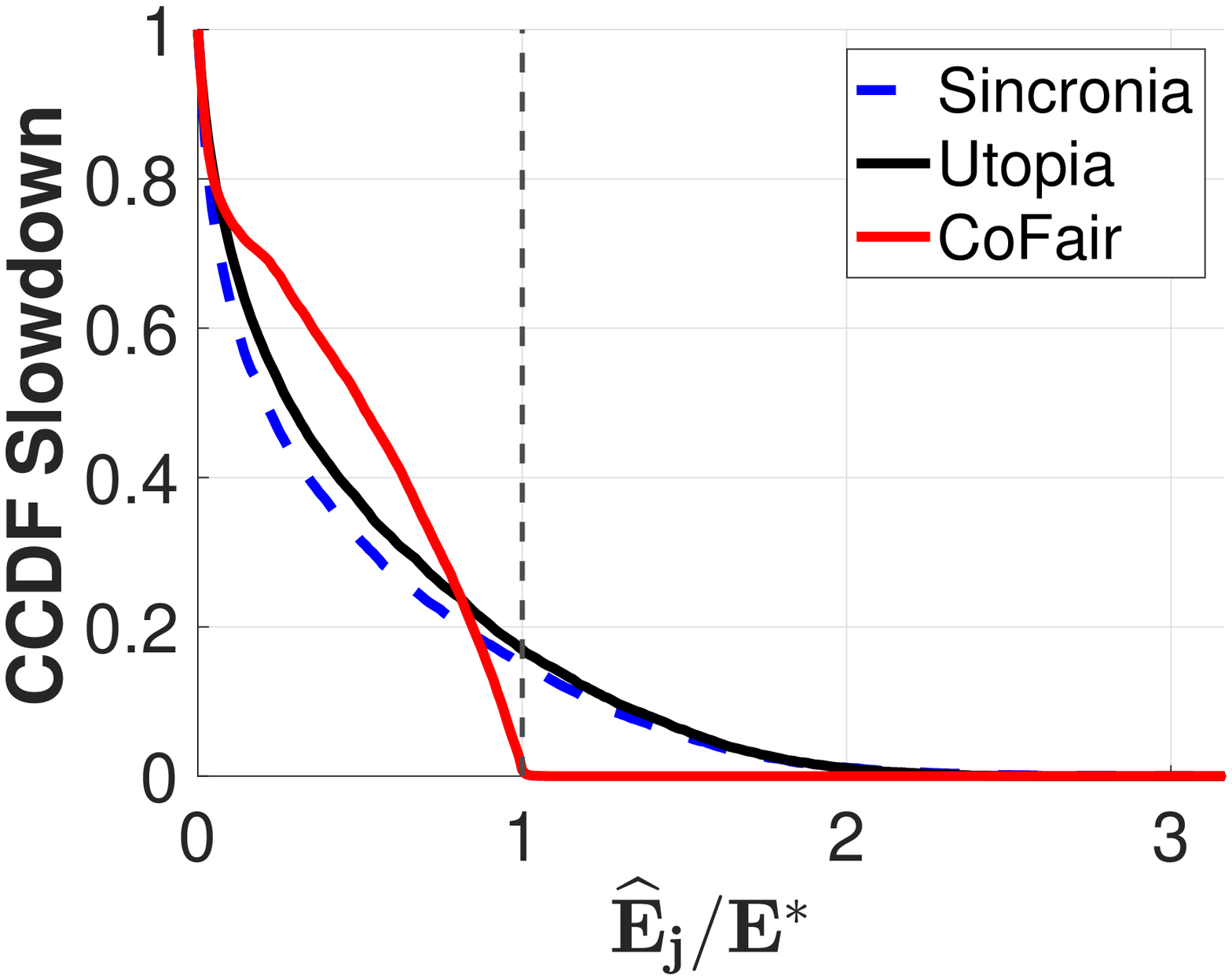} 
	\put(-125,0){c)}
	\end{minipage}
	\begin{minipage}{4.2cm}
	\includegraphics[scale=0.24]{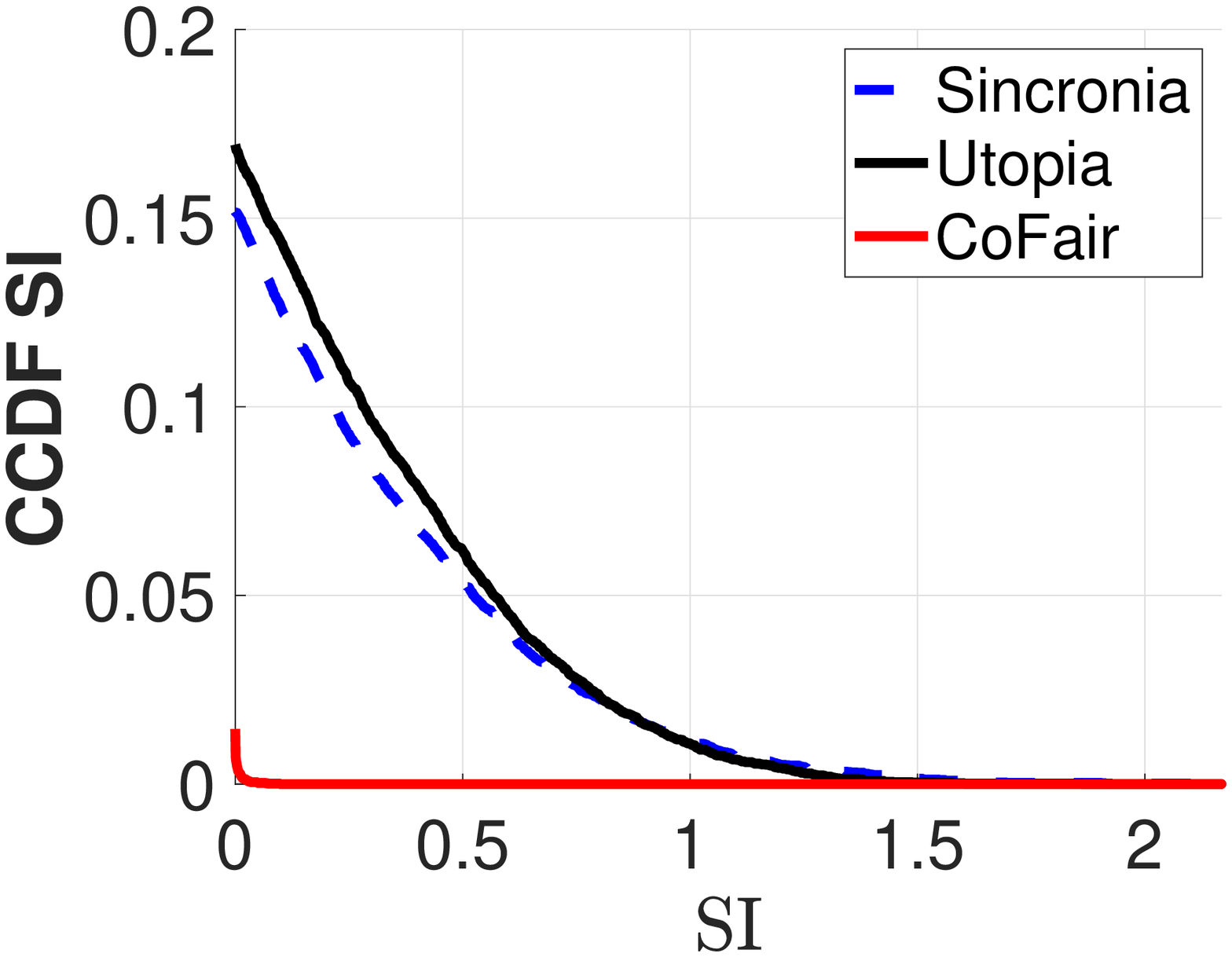} 
	\put(-125,0){d)}
	\end{minipage}
	\caption{MR traces, $\phi_j=V_j$: a) and c)  CCDF of the normalized slowdown; b) and d) CCDF of the stretch index; setting: $M=30$, and $N=30$, $m=10$ and $r=3$ for a) and b);$M=30$, and $N=100$, $m=10$ and $r=10$  $q=0.8$ and $N=100$ for c) and d).} \label{fig:num5}
\end{figure*}

\begin{figure*}[t]
        \centering
         \begin{minipage}{4.2cm}
	\includegraphics[scale=0.24]{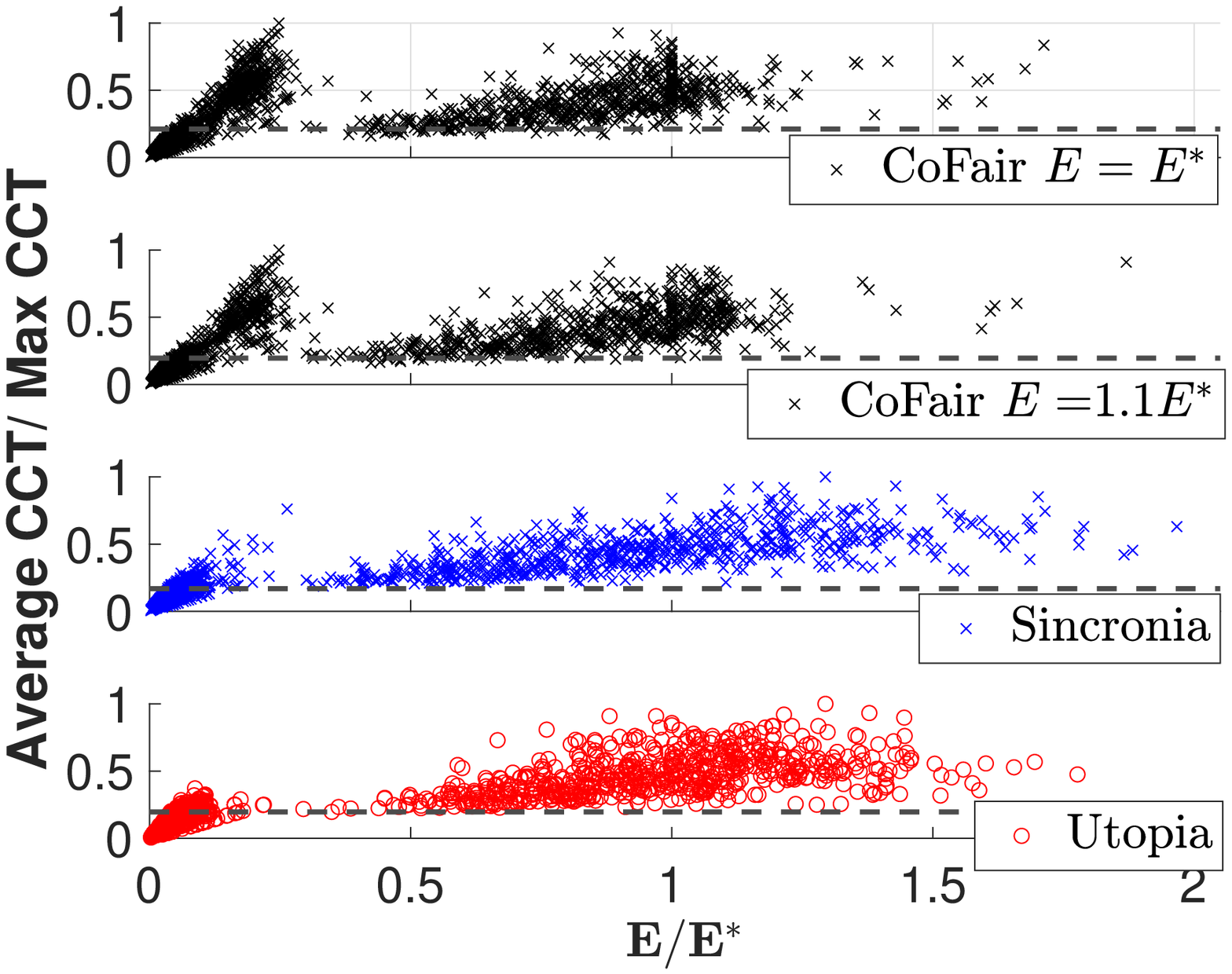} 
	\put(-125,0){a)}
	\end{minipage}
	\begin{minipage}{4.2cm}
	\includegraphics[scale=0.24]{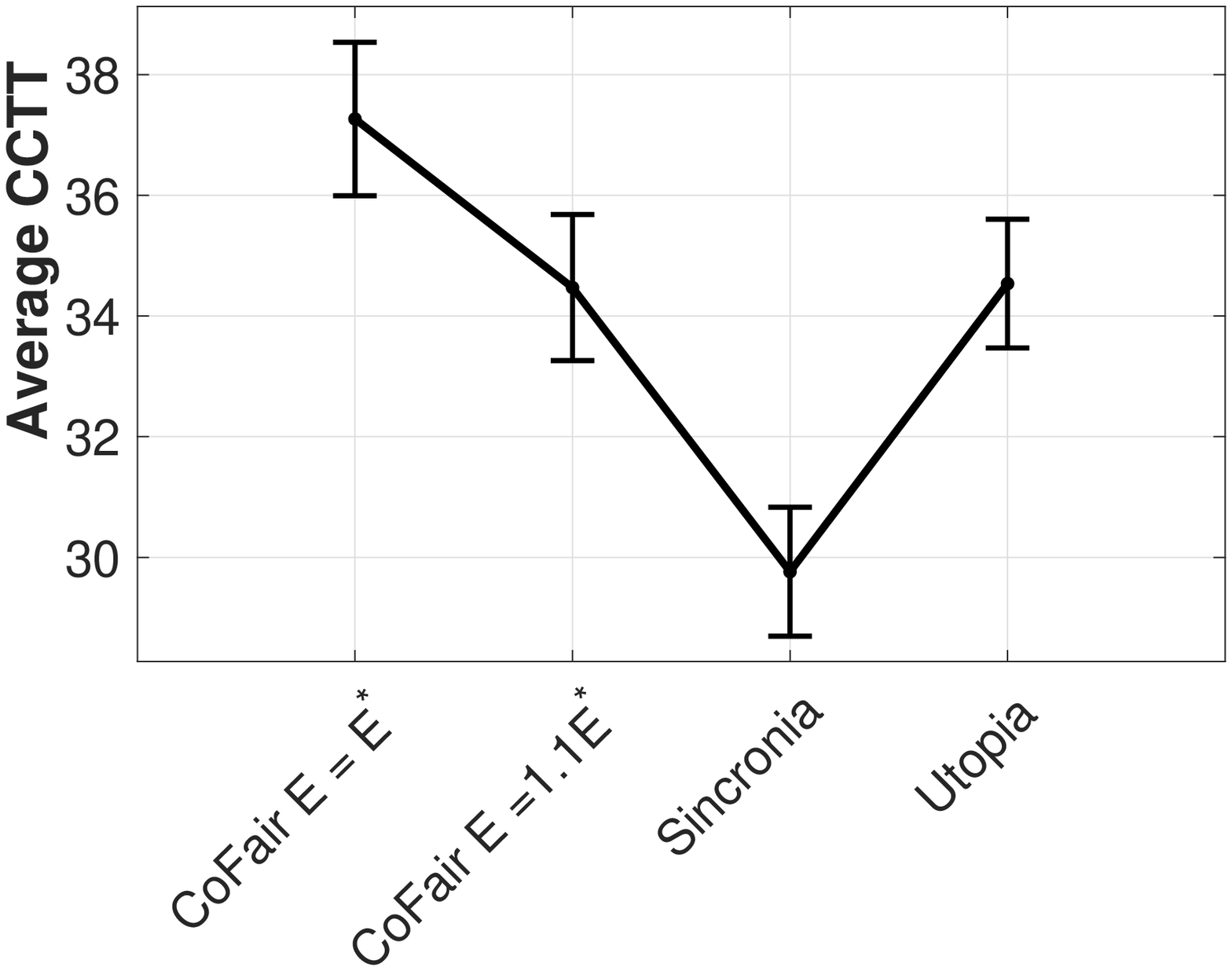} 
	\put(-125,0){b)}
	\end{minipage}
         \begin{minipage}{4.2cm}
	\includegraphics[scale=0.24]{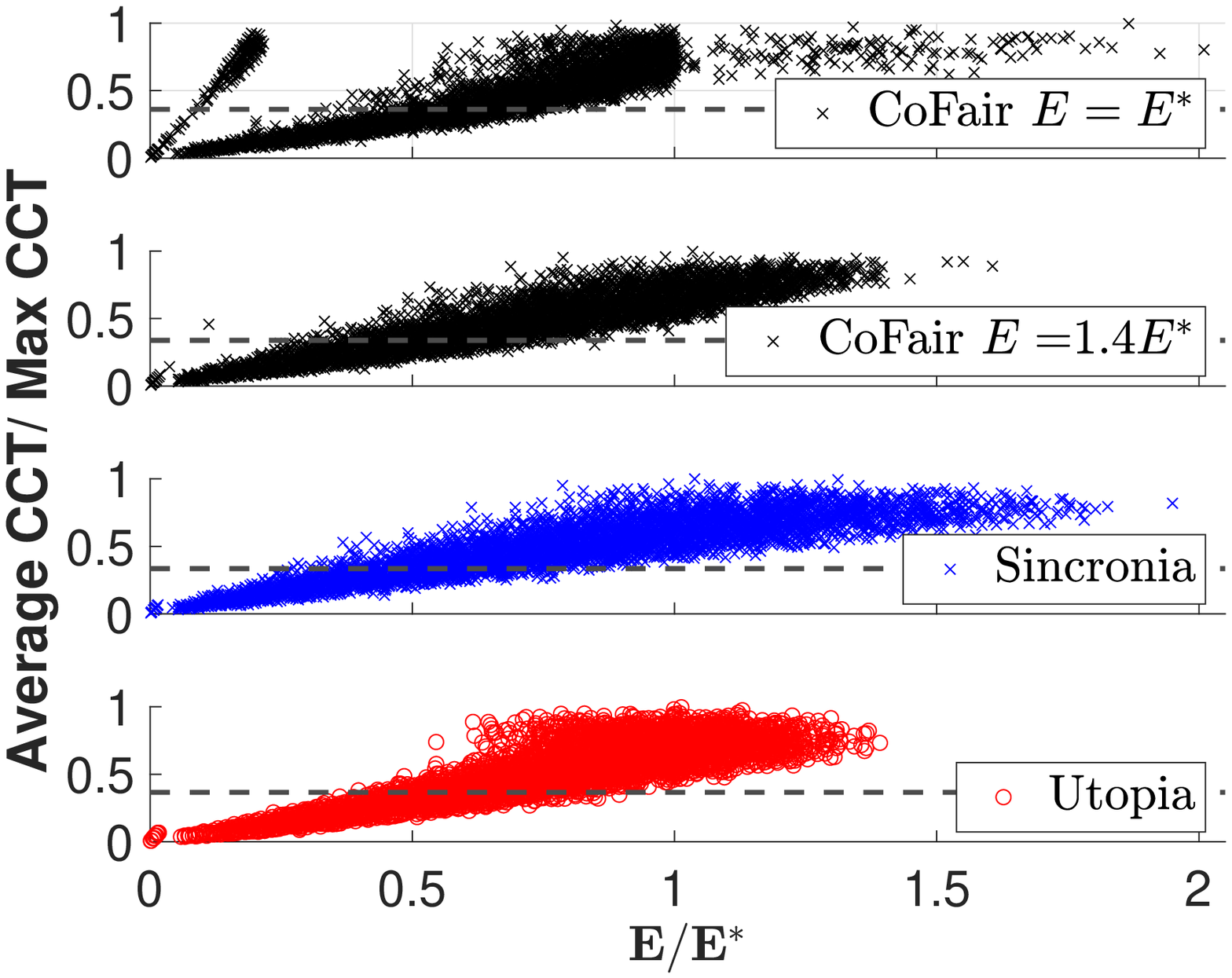} 
	\put(-125,0){c)}
	\end{minipage}
	\begin{minipage}{4.2cm}
	\includegraphics[scale=0.24]{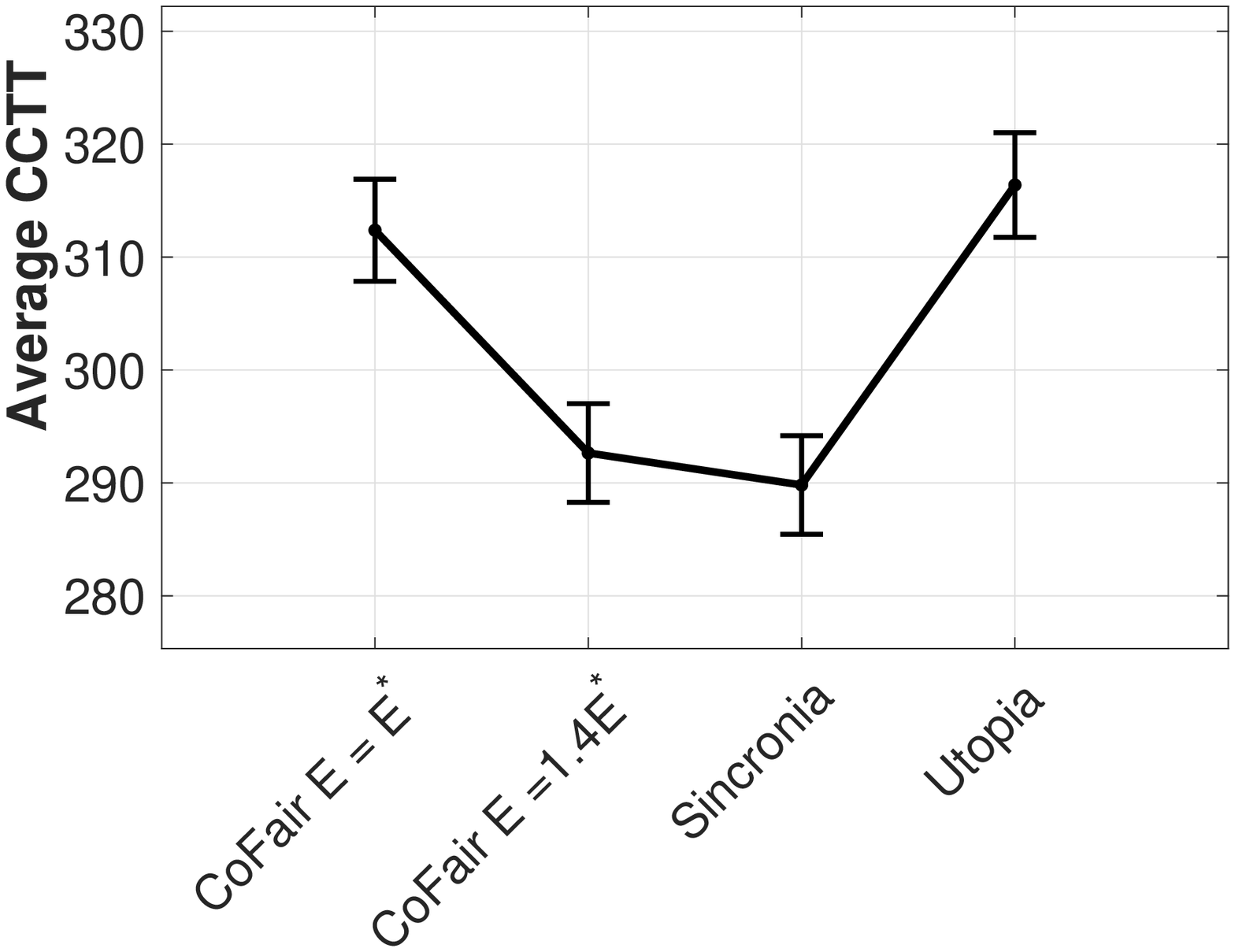} 
	\put(-125,0){d)}
	\end{minipage}
	\caption{WN traces, $\phi_j=V_j$:  a) and c) scatter diagram and b) and d) corresponding average CCT; setting:$M=30$, a) and b) $q=0.2$ and $N=30$ c) and d) $q=0.8$ and $N=100$.} \label{fig:num4bis}
\end{figure*}

\subsection{Slowdown with port occupation: $\phi_j=V_j$.} By letting $\phi_j=V_j$, the average port occupation is embedded in the slowdown definition: between two coflows with the same volume, it is fair if the one with lower average port occupation finishes first. Fig.~\ref{fig:num1tris}a and Fig.~\ref{fig:num1tris}b describe the effect of the slowdown constraint $\E$ onto the average CCT for a fabric with $M=30$ ports and $N=30,50,100$ coflows of type WN for $p=0.2$. Results are averaged on hundred sample batches and normalized against the average CCT of the near optimal solution provided by Sincronia; $95\%$ confidence intervals are superimposed. The average CCT decreases for increasing values of $\E$: this behavior mimics the $\alpha$ parameter of classic flow $\alpha$-fairness. As depicted in Fig.~\ref{fig:num1tris}a, $\E$ can be tuned in order to attain a tradeoff between coflow slowdown and average CCT, where the loss in CCT gain over Sincronia tops at $40\%$ for $p=0.2$ and $N=100$: this represents the loss in performance in order to grant all coflows the target slowdown. Same result is described in figure Fig.~\ref{fig:num1bis}d for $p=0.8$: in presence of many large coflows, the range of the tradeoff is smaller.  

Fig.~\ref{fig:num4}a/c and b/d report on the CCDF of the normalized slowdown and the corresponding stretch index, respectively, for two experiments on $100$ coflow batches. By comparing the distributions of the different algorithms for $p=0.2$, i.e., Fig.~\ref{fig:num2}a and Fig.~\ref{fig:num2}b, it is apparent that \cofair provides the best slowdown-efficiency tradeoff, whereas both baselines do not perform well both in terms of normalized slowdown and in terms of stretch index. The same behavior is observed for $p=0.8$: \cofair and Utopia perform similar in the inner interval, but the slowdown CCDF of \cofair bends clearly in proximity of  $E^*$ thus attaining a negligible number of violations. The tail of the SI distribution shows that their magnitude is relatively larger than those of Utopia in the same region. In these experiments, both Utopia and Sincronia have a rate of violations on the order of $10-20\%$.

The same experiment is repeated for the MR batches in Fig.~\ref{fig:num5}. As in the previous cases, Sincronia and \cofair provide the best performance in terms of slowdown. In this case, not only Sincronia  but also Utopia has a significant number of deviations, i.e., on the order of $10\%$ for coflows in the inner interval. \cofair provides the best tradeoff, outperforming Utopia in the outer interval. The experiment is hence performed for $m=10$ and $r=10$ and $N=100$ coflows. In this case, the CCDF tail of <\cofair in the outer interval is negligible, whereas Sincronia and Utopia show deviations larger than $15\%$.

Finally, in order to provide better understanding on the behavior of the algorithms,  Fig.~\ref{fig:num4bis}a and Fig.~\ref{fig:num4bis}b show the scatter diagram of the normalized CCT and the slowdown for each coflow corresponding to the experiments of Fig.~\ref{fig:num4}a and Fig.~\ref{fig:num4}b. As seen in Fig.~\ref{fig:num4bis}, in order to reduce the slowdown, \cofair tends to concentrate several coflows in the region where the slowdown is smaller at the price of higher CCT. In order to remark the tradeoff efficiency-fairness, the resulds for the runs of \cofair for $E=1.1 E^*$ have been added. As seen before in Fig.~\ref{fig:num2}, the performance in terms of slowdow is very similar to Utopia. However, here parameter $E$ acts as a fairness-efficiency knob as depicted in Fig.~\ref{fig:num5}b. In fact, while matching the same CCT of Utopia, \cofair attains a significantly smaller number of violations. Fig.~\ref{fig:num5}c and d report on the settings of Fig.~\ref{fig:num2}c and Fig.~\ref{fig:num2}d. Here, by letting on $E=1.4\, E^*$ \cofair overlaps the slowdown performance of Utopia, and yet it provides average CCT figures very close to that of Sincronia. 


\section{Conclusions and discussion.}\label{sec:conclusions}


The data transfer phase of modern computing frameworks requires to serve several concurrent coflows. This paper introduces a framework for coflow scheduling to trade off the average CCT for coflows' slowdown. It requires a single control parameter, that is the maximum slowdown for an input batch of coflows. Scheduling under slowdown contraints has larger feasibility set compared to the popular notion of progress based on minimum rate guarantees. The maximum possible progress, i.e., the minimum possible slowdown, can be calculated with great accuracy in $O(MN + N\log(N))$. 

A new scheduler, i.e., \cofair, has extended the framework of primal-dual $\sigma$-order schedulers to account for generalized slowdown constraints. The experimental results indicate that slowdown-aware $\sigma$-order generated by \cofair can adjust the priority of coflows with no apparent loss in performance with respect to near optimal benchmarks for average CCT minimization. Finally, based on the notion of generalized slowdown, \cofair and the proposed framework can perform coflow scheduling by accounting for specific features, e.g., the occupancy of the fabric ports.


\bibliographystyle{IEEEtran}
\bibliography{references.bib}

\appendix

\section*{Proof of Thm.~\ref{thm:feasib}}

\begin{IEEEproof}
The feasibility of the \ref{eq:dynsched} problem can be formulated with the following LP 
\begin{align}
	\mbox{minimize:}   \; 0  \label{sys:feas}\\
	\mbox{subj. to:} \;& \sum_{t=1}^T x_j^i(t)=1, \quad \forall j\in \C, i\in \F_j \label{eq:completion2}\\
     & Z_j(t) \leq 0,  \forall t \leq r_j \label{eq:release3}\\
     & Z_j(t) \geq 1,  \forall t >\frac{\E\cdot (C_j^0 - r_j)}{V_j} + r_j \label{eq:slowdown3}\\
	 & \sum_{j\in \C} \sum_{i \in \F_i} v_j^i x_j^i(t)\chi_j^i(\ell) \leq B_\ell, \;\forall t, \forall\ell \in \Ll \label{eq:cap}\\
     & x_j^i(t) \in [0,1],  \;\forall t 
\end{align}
The above LP is obtained by considering any rate allocation for which the constraints of \ref{eq:dynsched} are respected: \eqref{eq:slowdown3} has replaced the constraints \eqref{eq:CCT} and  \eqref{eq:elong} in \ref{eq:dynsched}. Since LP is in {\it P} this completes the proof of 
the first part of the statement. The second part statement follows by observing that from \eqref{eq:slowdown3}, $\frac{\E\cdot (C_j^0 - r_j)}{V_j} + r_j \geq  \frac{\E\cdot C_j^0 }{V_j} $ if and only if $\E\geq V_j$ for all $j\in \C$, which holds true since $C_j \geq C_j^0$ for all $j\in \C$. 
\end{IEEEproof}

\section*{Proof of Thm.~\ref{thm:feasible}}

\begin{IEEEproof}
Let define a feasible solution for the \ref{eq:primal} problem. For every pair $(\mu_k,\sigma(k))$  let define $C_{\sigma(k)}=\sum_{j\leq k}  p_{\mu_k,\sigma(j)}$; since \ref{eq:primal} is feasible, the slowdown constraint in the primal formulation is satisfied at each step as from Lemma~\ref{lem:feas}. Parallel inequalities hold by applying port-wise results in single machine scheduling \cite{queyranne}. 

For the \ref{eq:dual}, let consider solutions of the type  
\begin{equation}\label{eq:vars}
y_{\ell,A}=\begin{cases}
\theta_k >0 & \ell=\mu_k \;\mbox{and} \; A=F_k \; \mbox{for some} \; k\\
0 & \mbox{otherwise} 
\end{cases}
\end{equation}
where $\mu_k$ is the bottleneck selected at step $k$ and $\F_k$ is the corresponding set of tail-feasible coflows.   The corresponding form of the constraints of the dual problem
\begin{equation}\label{eq:sys2}
\sum_{k=j}^N q_{\mu_k,\sigma(j)} \,  y_{\mu_k,F_k} = w_{\sigma(j)} + \alpha_{\sigma(j)}, \quad j \in \C
\end{equation}
where $q_{\mu_k, j } = p_{\mu_k, j } \1\{j \in F_k \}$. If at each step $k=N,\ldots,2,1$ the selection is 
\begin{equation}\label{eq:choice}
\sigma(k) =\argmin\limits_{j \in F_k}\left \{\frac{ w_j^k}{p_{\mu_N,j}} \right \}
\end{equation} 
the algorithm produces a non-negative solution of the linear system \eqref{eq:sys2}. In fact, let first consider step $k=N$. The full rank system of the constraints of the dual system writes
\begin{equation}\label{eq:sysgauss}
\left (
 \begin{array}{cccc|c}
q_{_{\mu_N\sigma(N)}}  & 0 & \ldots & 0  & w_{\sigma(N)}^{(N+1)}\\
q_{_{\mu_N\sigma(N-1)}} & q_{_{\mu_{N-1}\sigma(N-1)}}  & \ldots & 0  &  w_{\sigma(N-1)}^{(N+1)}\\
\vdots  & \vdots  & \vdots  & \vdots  & \vdots\\
q_{_{\mu_N\sigma(1)}}&  q_{_{\mu_{N-1}\sigma(1)}} & \ldots & q_{_{\mu_1\sigma(1)}} &  w_{\sigma(1)}^{(N+1)}\\
    \end{array}
\right)\nonumber
\end{equation}
The Gaussian elimination using $\sigma(N)$ as pivot renders the new equivalent system 
\begin{equation}\label{eq:gauss}
\small
\left (\!\!
 \begin{array}{cccc|c}
1 & 0 & \ldots & 0  & \frac{w_{_{\sigma(N)}}^N}{p_{_{\mu_N,\sigma(N)}}}  \\[3mm]
0 &  p_{_{\mu_{N-1}\sigma(N-1)}}  & \ldots & 0  &  w_{_{\sigma(N-1)}}^N \!- \!w_{_{\sigma(N)}}^N\frac {p_{_{\mu_N\sigma(N-1)}} }{p_{_{\mu_N\sigma(N)}}}  
\\
\vdots  & \vdots  & \vdots  & \vdots  & \vdots\\
0 &  p_{_{\mu_{N-1}\sigma(1)}} & \ldots &  p_{_{\mu_1,\sigma(1)}}& w_{\sigma(1)}^N\!- \! w_{\sigma(N)}^N\frac {p_{_{\mu_N\sigma(1)}}}{p_{_{\mu_N\sigma(1)}}}   \\
\end{array}
\!\!\right )\nonumber
\end{equation}
Observe that the first column is nullified irrespective of the chosen permutation $\sigma(j)$ for $j<N$. The system has been transformed in the equivalent one where  Gaussian elimination can now be performed on the rightmost square sub-matrix. Let replace in the column on the right with the following values
\begin{eqnarray}
&&\small\hskip-7mm  w_j^{N-1}=\frac{  w_{\sigma(N)}^N}{p_{_{\mu_N\sigma(N)}}},  j=\sigma(N) \nonumber \\
&&\small\hskip-7mm  w_j^{N-1}=  w_{j}^N -   w_{\sigma(N)}^N\frac {p_{\mu_N,j} }{p_{\mu_N,\sigma(N)}}\geq 0,  j\in F_k \setminus\{ \sigma(N) \}\nonumber\\
&&\small\hskip-7mm  w_j^{N-1}=  w_{j}^N,  j\not \in F_k\nonumber
\end{eqnarray}
where $w_j^{N-1}\geq 0$ for all $j\in \C$. Indeed, $y_{\mu_N F_N}=\frac{  w_{\sigma(N)}^N}{p_{_{\mu_N\sigma(N)}}}$. The algorithm iterates the procedure on the remaining subsystem using the corner element of $\sigma(k)$  as pivot so that at each step $y_{\mu_k,F_k}=\theta_k=\frac{  w_{\sigma(k)}^N}{p_{_{\mu_k\sigma(k)}}}\geq 0$.
\end{IEEEproof}
Note that unscheduled coflows which are non primal-feasible at a given step do not take part to gaussian elimination. 

\section*{Proof of Thm.~\ref{thm:completeness}}
\begin{IEEEproof}
The sketch of the proof is as follows. First, by linearity rescaling weights $\{\kappa \cdot w_j\}$ is irrelevant. The Smith ratio at the $k$-th step for a candidate coflow 
\begin{equation*}
\begin{small}
\!\frac{w_{_{\sigma(N-k)}}^{^{(N-k)}}}{P_{_{\mu_k\sigma(N-k)}}}\!=\!\frac{w_{_{\sigma(N-k)}}\!\!\!+\alpha_{_{\sigma(N-k)}}\!\!\!-\sum_{h=1}^k w_{_{\sigma(N-k+h)}}^{^{(N-k+h)}}\frac{P_{_{\mu_h\sigma(N-k)}}}{P_{_{\mu_h\sigma(N-k+h)}}}}{P_{_{\mu_k\sigma(N-k)}}}
\end{small}
\end{equation*}
Hence, to enforce the desired $\sigma$-order $\sigma^*$ as output, let start by an input set $\alpha_j=1$, for all $j$, and choose $\alpha_{\sigma^*(N-k)}$ such in a way that the Smith ratio of for $\sigma^*(N-k)$ be minimal among all candidate coflows, simply by rendering $(w_{_{\sigma(N-k)}}+\alpha_{_{\sigma(N-k)}})$ small enough as compared to the other coflows, possibly rescaling all weights $w_j$ by $\kappa$. Using $\{\alpha_j\}$ as input to the algorithm produces the desired output $\sigma^*$.
\end{IEEEproof}

\section*{Proof of Thm.~\ref{lma:primal}}

\begin{IEEEproof}
For $C_j,$ output of Alg.~\ref{alg:greedy}, it holds
\begin{small}
\begin{eqnarray}
&& \sum_{j= 1}^N w_j C_j = \sum_{j= 1}^N C_j  \Big ( \sum_{\ell \in \Ll} \sum_{A \subseteq \C : j \in S}  y_{\ell,A} \, p_{\ell,j}  - \alpha_j \Big ) \nonumber \\
&& = \sum_{j= 1}^N C_{\sigma(j)} \Big ( \sum_{k=j}^N  p_{\mu_k,\sigma(j)} y_{\mu F_k} - \alpha_{\sigma(j)} \Big ) \nonumber \\
&& \stackrel{(i)}{\leq} \sum_{k = 1}^N y_{\mu_kF_k}  \sum_{j=1}^k  p_{\mu_k\sigma(j)} C_{\sigma(j)}  -  \sum_{j=1}^N \alpha_j C_j  \nonumber \\
&& \stackrel{(ii)}{\leq} \sum_{k = 1}^N y_{\mu_kF_k}  \Big ( \sum_{j=1}^k  p_{\mu_k\sigma(j)} \Big )^2  -  \sum_{j=1}^N \alpha_j C_j  \nonumber \\
&& \leq  2 \Big ( \sum_{k = 1}^N y_{\mu_kF_k} f_{\mu_k}(F_k)  - \alpha_k C_k \Big )   + \sum_{j=1}^N \alpha_j (2 D_j - C_j )   \nonumber \\
&& \stackrel{(iii)}{\leq}  2  \sum_{k = 1}^N w_k  \Clp{k}  + \sum_{j=1}^N \alpha_j (2 D_j - C_j )   \nonumber \\
&& \leq  2  \sum_{k = 1}^N w_k  \Copt{k}  + 2 \sum_{j=1}^N \alpha_j D_j    \nonumber 
\end{eqnarray}
\end{small}
Where the following arguments hold: $(i)$ follows from the fact that in general $j\in F_k$ does not imply $j\in F_{k+1}$ and $(ii)$ follows from the fact that $C_{\sigma(k)} = \sum_{j=1}^k  p_{\mu_k\sigma(j)} \geq C_{\sigma(j)}$ for all $k=1,\ldots,j$. According to Lemma.~\ref{lma:results}, it holds 
\[
\sum_{j= 1}^N w_j  \Csig{j}   \leq 2  \sum_{j= 1}^N w_j C_j  \leq 4  \sum_{k = 1}^N w_k  \Copt{k}  + 4 \sum_{j=1}^N \alpha_j D_j 
\]
which concludes the proof.  
\end{IEEEproof}

\end{document}